\documentclass[11pt,fleqn]{article}
\usepackage{amssymb,latexsym,amsmath,amsfonts,graphicx,curves,ebezier}

    \advance\textheight by \topskip

    \numberwithin{equation}{section}

    \def\Re{{\rm Re \,}}
    
    \def\Ai{{\rm Ai \,}}

    \def\bigO{{\cal O}}
    \def\Res{{\rm Res \,}}

    \newtheorem{theorem}{Theorem}[section]
    \newtheorem{lemma}[theorem]{Lemma}
    
    \newtheorem{proposition}[theorem]{Proposition}
    \newtheorem{conjecture}[theorem]{Conjecture}
    \newtheorem{Definition}[theorem]{Definition}
    
    \newtheorem{Remark}[theorem]{Remark}
    \newenvironment{remark}{\begin{Remark}\rm}{\end{Remark}}
    \newtheorem{Example}[theorem]{Example}

    \newenvironment{proof}%
    {\rm \trivlist \item[\hskip \labelsep{\bf Proof. }]}%
    {\hspace*{\fill}$\Box$\endtrivlist}
    \newenvironment{varproof}%
    {\rm \trivlist \item[\hskip \labelsep{\bf Proof}]}%
    {\hspace*{\fill}$\Box$\endtrivlist}

    \newcommand{\sgn}{{\operatorname{sgn}}}
    \newcommand{\Arg}{{\operatorname{Arg}}}

\begin{document}

\title{The existence of a real pole-free solution of the fourth order
    analogue of the Painlev\'e I equation}

\author{T. Claeys and M. Vanlessen}

\maketitle

\begin{abstract}
    We establish the existence of a real solution $y(x,T)$ with no
    poles on the real line of the following fourth order analogue of
    the Painlev\'e I equation,
    \[
        x=Ty-\left(\frac{1}{6}y^3+\frac{1}{24}(y_x^2+2yy_{xx})
            +\frac{1}{240}y_{xxxx}\right).
    \]
    This proves the existence part of a conjecture posed by Dubrovin.
    We obtain our result by proving the solvability of an associated
    Riemann-Hilbert problem through the approach of a vanishing lemma.
    In addition, by applying the Deift/Zhou steepest-descent method to
    this Riemann-Hilbert problem, we obtain the asymptotics for
    $y(x,T)$ as $x\to\pm\infty$.
\end{abstract}

\section{Introduction}

\subsection{The $P_I^2$ equation}

The first Painlev\'e equation is the second order differential
equation
\begin{equation}
    y_{xx}=6y^2+x.
\end{equation}
This equation has higher order analogues of even order $2m$ for
$m\geq 1$, which are collected, together with the first Painlev\'e
equation itself, in the Painlev\'e I hierarchy, see
e.g.~\cite{KKNT,KS}. The second member in the hierarchy is the
fourth order differential equation
\begin{equation}\label{PI2withoutT}
    x=-\left(\frac{1}{6}y^3+\frac{1}{24}(y_x^2+2yy_{xx})
        +\frac{1}{240}y_{xxxx}\right),
\end{equation}
and has solutions that are meromorphic in the complex plane. In
1990, Br\'ezin, Marinari, and Parisi \cite{BMP} argued numerically
that there exists a solution $y$ to (\ref{PI2withoutT}) with no
poles on the real line, and with asymptotic behavior
\begin{equation}\label{asymptotics y}
    y(x)\sim \mp|6x|^{1/3},
        \qquad\mbox{as $x\to\pm\infty$.}
\end{equation}
Moore \cite{Moore} proved the existence of a unique real solution
to (\ref{PI2withoutT}) with asymptotic behavior given by
(\ref{asymptotics y}), and he gave a line of argument why this
solution is probably pole-free on the real line.

A generalization of (\ref{PI2withoutT}) can be obtained by
introducing an additional variable $T$, as done by Dubrovin in
\cite{Dubrovin}, so that we get the following differential equation
for $y=y(x,T)$, which we denote as the $P_I^2$ equation
(cf.~\cite{Kapaev} for $T=0$),
\begin{equation}\label{PI2}
    x=Ty-\left(\frac{1}{6}y^3+\frac{1}{24}(y_x^2+2yy_{xx})
        +\frac{1}{240}y_{xxxx}\right).
\end{equation}
In recent work \cite{Dubrovin}, Dubrovin conjectured (see Section
\ref{subsection: motivation} below for more details) the existence
of a unique real solution to (\ref{PI2}) with no poles on the real
line. We prove the existence part of this conjecture.

Our result is the following.
\begin{theorem}\label{main theorem}
    There exists a solution $y(x,T)$ to the $P_I^2$ equation {\rm
    (\ref{PI2})} with the following properties:
    \begin{itemize}
    \item[(i)]  $y(x,T)$ is real valued and pole-free
    for $x,T\in\mathbb{R}$.
    \item[(ii)] For fixed $T\in\mathbb{R}$, $y(x,T)$ has the following
    asymptotic behavior,
    \begin{equation}\label{main theorem: eq1}
        y(x,T)=\frac{1}{2}z_0 |x|^{1/3}+\bigO(|x|^{-2}),
            \qquad\mbox{as $x\to\pm\infty$,}
    \end{equation}
    where $z_0=z_0(x,T)$ is the real solution of
    \begin{equation}
        z_0^3=-48\sgn(x)+24z_0|x|^{-2/3}T.
    \end{equation}
    \end{itemize}
\end{theorem}

\begin{remark}\label{remark: asymptotics z0}
    Observe that $z_0$ is negative (positive) for $x>0$ ($x<0$) with
    the following asymptotic behavior as $x\to\pm\infty$,
    \begin{equation}\label{asymptotics z0}
        z_0=\hat z_0-\sgn(x)\,\frac{2}{3}\,6^{2/3}T|x|^{-2/3}+
            \bigO(|x|^{-4/3}),
            \qquad \hat z_0=-\sgn(x)\,2\cdot 6^{1/3},
    \end{equation}
    so that the asymptotics (\ref{main theorem: eq1}) for $y$ can be
    rewritten as, cf.~(\ref{asymptotics y})
    \begin{equation}
        y(x,T)=\mp (6|x|)^{1/3}\mp \frac{1}{3}6^{2/3}T|x|^{-1/3}
            +\bigO(|x|^{-1}),
            \qquad\mbox{as $x\to\pm\infty$.}
\end{equation}
    Power expansions for solutions of (\ref{PI2withoutT}) were found
    in \cite{KE}.
\end{remark}

\begin{remark}
    One expects, see \cite[Appendix A]{Moore} for $T=0$, that the
    solution $y$ considered in Theorem \ref{main theorem} is
    uniquely determined by realness and the asymptotics
    (\ref{main theorem: eq1}).
\end{remark}

\subsection{Motivation}
\label{subsection: motivation}

\subsubsection*{Hamiltonian perturbations of hyperbolic equations}

Hyperbolic equations of the form
\begin{equation}\label{hyperbolic equation}
    u_t+a(u)u_x=0
\end{equation}
can be perturbed to a Hamiltonian equation of the form
\begin{multline}\label{hamiltonian perturbation}
    u_t+a(u)u_x+\epsilon\left[b_1(u)u_{xx}+b_2(u)u_x^2\right]
    \\[1ex]
    +\epsilon^2\left[b_3(u)u_{xxx}+b_4(u)u_xu_{xx}+b_5(u)u_x^3\right]+\cdots =0,
\end{multline}
where $\epsilon$ is small and $b_1, b_2, \ldots$ are smooth
functions. These equations have been studied by Dubrovin in
\cite{Dubrovin}, see also \cite{DubrovinI}, where he formulated the
universality conjecture about the behavior of a generic solution to
a general perturbed Hamiltonian equation (\ref{hamiltonian
perturbation}) near the point $(x_0,t_0)$ of {\em gradient
catastrophe} of the unperturbed solution (\ref{hyperbolic
equation}). He argued that this behavior is described by a special
solution to the $P_I^2$ equation (\ref{PI2}). To be more precise,
his conjecture is the following.

\begin{conjecture}\label{conjecture: Dubrovin}{\rm (Dubrovin,
\cite{Dubrovin})}
\begin{itemize}
\item[(i)]Let $u_0=u_0(x,t)$ be a smooth solution to the
unperturbed equation {\rm (\ref{hyperbolic equation})}, defined for
all $x\in\mathbb R$ and $0\leq t < t_0$, and monotone in $x$ for any
$t$. Then there exists a solution $u=u(x,t;\epsilon)$ to the
perturbed equation {\rm (\ref{hamiltonian perturbation})} defined on
the same domain in the $(x,t)$-plane with the asymptotics as
$\epsilon\to 0$ of the form
\begin{equation}
u(x,t;\epsilon)=u_0(x,t)+\epsilon^2 u_1(x,t)+\epsilon^4
u_2(x,t)+o(\epsilon^4),
\end{equation}
where $u_1$ and $u_2$ can be written down explicitly. \item[(ii)]
The ODE {\rm (\ref{PI2})} has a unique solution $y=y(x,T)$ smooth
for all real $x\in\mathbb R$ for all values of the parameter $T$.
\item[(iii)]The generic solution $u$ described in part (i) of the
conjecture can be extended up to $t=t_0+\delta$ for sufficiently
small positive $\delta=\delta(\epsilon)$; near the point $(x_0,t_0)$
it behaves in the following way
\[u(x,t;\epsilon)=u_0(x,t)+a\epsilon^{2/7}
y\left(b\epsilon^{-6/7}(x-c(t-t_0)-x_0),d\epsilon^{-4/7}(t-t_0)\right)
+\bigO(\epsilon^{4/7}),
\]
for some constants $a$, $b$, $c$, $d$ which depend on the hyperbolic
equation, the solution $u$, and on the choice of perturbation. Here
$y$ is the unique smooth solution described in part (ii) of the
conjecture.
\end{itemize}
\end{conjecture}

So Theorem \ref{main theorem} in fact proves the existence part of
part (ii) of Dubrovin's conjecture.

\medskip

In \cite{GK}, numerical calculations were done for the particular
example (of a perturbed Hamiltonian equation) of the small
dispersion limit of the KdV equation, see also \cite{LL1, LL2, LL3,
Venakides},
\[u_t+6uu_x+\epsilon^2u_{xxx}=0, \qquad \textrm{ with initial condition }\quad
u(x,0)=u_0(x).\] Before the time of gradient catastrophe $t_0$,
solutions turn out to behave nicely. When approaching the critical
time $t_0$, the slope of the function blows up near $x_0$, and at
the critical time, fast oscillations near $x_0$ set in. The
transition between the monotone behavior and the oscillations should
be described in terms of the real pole-free solution to (\ref{PI2})
we consider in this paper.

\subsubsection*{Random matrix theory}

The local eigenvalue correlations of unitary random matrix ensembles
on the space of $n\times n$ Hermitian matrices have universal
behavior (when the size $n$ of the matrices is going to infinity) in
different regimes of the spectrum. In the bulk of the spectrum it is
known, see e.g.~\cite{BI1,Deift,DKMVZ2,PS}, that the local
correlations can be expressed in terms of the sine kernel, while at
the soft edge of the spectrum they generically (i.e.~when the
limiting mean eigenvalue density vanishes like a square root) can be
expressed in terms of the Airy kernel, see
e.g.~\cite{BI1,DKMVZ2,Forrester,TracyWidom}.

In the presence of singular points, one observes different types of
limiting kernels in double scaling limits, see e.g.~\cite{BI, CK,
CKV}. Near singular edge points, where the limiting mean eigenvalue
density vanishes at a higher order than a square root (the regular
case) the local eigenvalue correlations are expected \cite{BB} to be
described in terms of functions associated with real pole-free
solutions of the even members of the Painlev\'e I hierarchy. The
particular case where the limiting mean eigenvalue density vanishes
like a power $5/2$, which is the lowest non-regular order of
vanishing, should correspond with the real pole-free solution of
$P_I^2$ considered in Theorem \ref{main theorem}. We intend to come
back to this in a future publication.

%

\subsection{Riemann-Hilbert problem and Lax pair for $P_I^2$}\label{subsection: lax pair}
Consider the following Riemann-Hilbert (RH) problem for given
complex parameters $x$ and $T$, on a contour
$\Sigma=\bigl(\cup_{j=0}^6\Sigma_j\bigr)\cup \mathbb R^-$, with
$\Sigma_j=e^{j\frac{2\pi i}{7}}\mathbb R^+$, where each of the
eight rays are orientated from $0$ to infinity.

\subsubsection*{RH problem for $\Psi$:}

\begin{itemize}
    \item[(a)] $\Psi$ is analytic in $\mathbb{C}\setminus\Sigma$.
    \item[(b)] $\Psi$ satisfies the following jump relations on
    $\Sigma$, for some complex numbers $s_0, \ldots, s_6$ which do
    not depend on $\zeta$, $x$, and $T$,
    \begin{align}
        \Psi_+(\zeta)&=\Psi_-(\zeta)
            \begin{pmatrix}
                1 & s_j \\
                0 & 1
            \end{pmatrix},&& \mbox{for $\zeta\in\Sigma_j$ for even $j$,}\\[1ex]
        \Psi_+(\zeta)&=\Psi_-(\zeta)
            \begin{pmatrix}
                1 & 0 \\
                s_j & 1
            \end{pmatrix},&& \mbox{for
            $\zeta\in\Sigma_j$ for odd $j$,}\\[1ex]
        \Psi_+(\zeta)&=\Psi_-(\zeta)
            \begin{pmatrix}
                0 & -1 \\
                1 & 0
            \end{pmatrix},&& \mbox{for $\zeta\in\mathbb R^-$.}
    \end{align}
    \item[(c)] There exist complex numbers $y$ and $h$, which depend on $x$ and $T$ but not
    on $\zeta$, such that $\Psi$ has the following asymptotic behavior as
    $\zeta\to\infty$,
    \begin{equation}
        \Psi(\zeta)=\zeta^{-\frac{1}{4}\sigma_3}N\left(I-h\sigma_3\zeta^{-1/2}
        +\frac{1}{2}\begin{pmatrix}h^2 & iy\\-iy &
        h^2\end{pmatrix}\zeta^{-1}+\bigO(\zeta^{-2})\right)
        e^{-\theta(\zeta;x,T)\sigma_3},
    \end{equation}
    where
    \begin{equation}\label{definition: N theta}
        N=\frac{1}{\sqrt 2}
        \begin{pmatrix}
             1 & 1 \\
             -1 & 1
        \end{pmatrix}e^{-\frac{1}{4}\pi i\sigma_3},\qquad
        \theta(\zeta;x,T)=\frac{1}{105}\zeta^{7/2}-\frac{1}{3}T\zeta^{3/2}+x\zeta^{1/2}.
    \end{equation}
\end{itemize}

\begin{remark}
In \cite{Kapaev}, Kapaev uses a slightly modified RH problem for
the $P_I^2$ equation with parameter $T=0$.  However a
transformation shows that both RH problems are equivalent.
\end{remark}
\begin{remark}\label{remark: RHPP1hierarchy}
The RH problem for $P_I^2$ is similar to the RH problem for the
Painlev\'e I equation, see \cite{Kapaev2}. The only differences
are that, for Painlev\'e I, there are only six rays in the jump
contour, and that the highest exponent of $\zeta$ in $\theta$ is
$5/2$. For the $m$-th member of the Painlev\'e I hierarchy, there
are $4+2m$ rays in the jump contour, and the highest exponent of
$\zeta$ in $\theta$ is $m+3/2$.
\end{remark}

The complex numbers $s_0, \ldots, s_6$ are the Stokes multipliers
and do not depend on $x$ and $T$, so that varying the parameters
$x$ and $T$ leads to a monodromy preserving deformation \cite{FN,
GP, IN, JimboMiwa}. The RH problem can only be solvable if the
Stokes multipliers satisfy the relation
\begin{equation}\label{condition stokes}
\begin{pmatrix}1&0\\s_4&1\end{pmatrix}
\begin{pmatrix}1&s_5\\0&1\end{pmatrix}
\begin{pmatrix}1&0\\s_6&1\end{pmatrix}
\begin{pmatrix}1&s_0\\0&1\end{pmatrix}
\begin{pmatrix}1&0\\s_1&1\end{pmatrix}
\begin{pmatrix}1&s_2\\0&1\end{pmatrix}
\begin{pmatrix}1&0\\s_3&1\end{pmatrix}
=\begin{pmatrix}0&1\\-1&0\end{pmatrix}.
\end{equation}
As we will show in Section \ref{subsection: 2.3} (in fact we only
treat one particular choice of Stokes multipliers, but the proof
holds in general), a solution $\Psi$ of the RH problem for $\Psi$
also satisfies the following system of differential equations, which
is the Lax pair for the $P_I^2$ equation,
\begin{equation}\label{differential equations}
    \frac{\partial \Psi}{\partial\zeta}=U \Psi,\qquad
    \frac{\partial \Psi}{\partial x}=W \Psi,
\end{equation}
where
\begin{align}\label{introduction: U}
    & U = \frac{1}{240}
    \begin{pmatrix}
        -4y_x \zeta-(12yy_x+y_{xxx}) &
        8\zeta^2+8y\zeta+(12y^2+2y_{xx}-120T) \\[1ex]
        U_{21}
        & 4y_x \zeta+(12yy_x+y_{xxx})
    \end{pmatrix},\\[3ex]
    & U_{21} =
    8\zeta^3-8y\zeta^2-(4y^2+2y_{xx}+120T)\zeta+
    (16y^3-2y_x^2+4yy_{xx}+240x),
\end{align}
and
\begin{equation}\label{introduction: W}
    W =\begin{pmatrix}
        0 & 1 \\
        \zeta-2y & 0
    \end{pmatrix}.
\end{equation}

The compatibility condition of the Lax pair (\ref{differential
equations})--(\ref{introduction: W}) is exactly the $P_I^2$
equation (\ref{PI2}), see e.g.~\cite{KKNT} for $T=0$. Different
choices of Stokes multipliers $s_0, \ldots, s_6$ correspond to
different solutions of the $P_I^2$ equation. The particular
solution we are interested in, is the unique solution with Stokes
multipliers $s_1=s_2=s_5=s_6=0$. It then follows by
(\ref{condition stokes}) that $s_0=1$ and $s_3=s_4=-1$. Kapaev
conjectured in \cite{Kapaev} that, for $T=0$, this solution has
asymptotics for $x\to\pm\infty$ which agree with (\ref{main
theorem: eq1}).

\subsection{Outline of the rest of the paper}

In the next section, we prove the first part (the existence part)
of Theorem \ref{main theorem}. In order to do this, we introduce
in Section \ref{subsection: RH problem for Phi} a RH problem for
$\Phi$, which is equivalent to the RH problem for $\Psi$ (the RH
problem for $P_I^2$) with Stokes multipliers $s_1=s_2=s_5=s_6=0$,
$s_0=1$, and $s_3=s_4=-1$. Afterwards, we prove in Section
\ref{subsection: solvability RH problem for Phi} the solvability
of the RH problem for $\Phi$ for real $x$ and $T$ by proving that
the associated homogeneous RH problem has only the trivial
solution. This approach is often referred to in the literature as
a vanishing lemma, see
e.g.~\cite{CKV,DKMVZ2,FokasMuganZhou,FokasZhou,Zhou}. We are only
able to prove the vanishing lemma for real $x$ and $T$ due to
symmetries in the RH problem. In Section \ref{subsection: 2.3} we
show that $\Psi$ satisfies a Lax pair of the form
(\ref{differential equations})--(\ref{introduction: W}), with $y$
given in terms of $\Phi$. By compatibility of the Lax pair, it
follows that $y$ solves the $P_I^2$ equation, and by the
solvability of the RH problem, $y$ has no real poles.

In Section $3$ we prove the second part (the asymptotics part) of
Theorem \ref{main theorem}. We do this by applying the Deift/Zhou
steepest-descent method \cite{DZ1,DZ2} to the RH problem for $\Phi$.
In this method, we perform a series of transformations to reduce the
RH problem for $\Phi$ to a RH problem that we can solve
approximately for large $|x|$. By unfolding the series of the
transformations, we obtain the asymptotics for $y$.

\section{The existence of a real pole-free solution to $P_I^2$}

\subsection{Statement of an associated RH problem to $P_I^2$}
\label{subsection: RH problem for Phi}

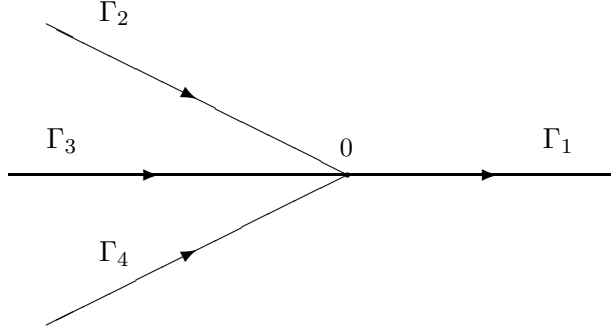
\begin{figure}[t]
\begin{center}
    \setlength{\unitlength}{1truemm}
    \begin{picture}(100,48.5)(0,2.5)
        \put(50,27.5){\thicklines\circle*{.8}}
        \put(49,30){\small 0}

        \put(50,27.5){\line(-2,1){40}}
        \put(50,27.5){\line(-2,-1){40}}
        \put(5,27.5){\line(1,0){80}}

        \put(76,31){$\Gamma_1$}
        \put(17,48){$\Gamma_2$}
        \put(10,31){$\Gamma_3$}
        \put(17,16){$\Gamma_4$}

        \put(30,37.5){\thicklines\vector(2,-1){.0001}}
        \put(30,17.5){\thicklines\vector(2,1){.0001}}
        \put(25,27.5){\thicklines\vector(1,0){.0001}}
        \put(70,27.5){\thicklines\vector(1,0){.0001}}
    \end{picture}
    \caption{The oriented contour $\Gamma$ consisting of the four straight rays
    $\Gamma_1$, $\Gamma_2$, $\Gamma_3$, and $\Gamma_4$.}
    \label{figure: RHP Psi}
\end{center}
\end{figure}

Let $\Gamma=\bigcup_{j=1}^4\Gamma_j$ be the contour consisting of
four straight rays,
\[
    \Gamma_1:\arg\zeta=0,\qquad\Gamma_2:\arg\zeta=\frac{6\pi}{7},\qquad\Gamma_3:\arg\zeta=\pi,
    \qquad\Gamma_4:\arg\zeta=-\frac{6\pi}{7},
\]
oriented as shown in Figure \ref{figure: RHP Psi}. We seek (for
$x,T\in\mathbb{C}$) a $2\times 2$ matrix valued function
$\Phi(\zeta;x,T)=\Phi(\zeta)$ (we suppress notation of $x$ and $T$
for brevity) satisfying the following RH problem.

\subsubsection*{RH problem for $\Phi$:}

\begin{itemize}
    \item[(a)] $\Phi$ is analytic in $\mathbb{C}\setminus\Gamma$.
    \item[(b)] $\Phi$ satisfies the following constant jump relations on
    $\Gamma$,
    \begin{align}
        \label{RHP Phi: b1}
        \Phi_+(\zeta)&=\Phi_-(\zeta)
            \begin{pmatrix}
                1 & 1 \\
                0 & 1
            \end{pmatrix},& \mbox{for $\zeta\in\Gamma_1$,}\\[1ex]
        \label{RHP Phi: b2}
        \Phi_+(\zeta)&=\Phi_-(\zeta)
            \begin{pmatrix}
                1 & 0 \\
                1 & 1
            \end{pmatrix},& \mbox{for
            $\zeta\in\Gamma_2\cup\Gamma_4$,}\\[1ex]
        \label{RHP Phi: b3}
        \Phi_+(\zeta)&=\Phi_-(\zeta)
            \begin{pmatrix}
                0 & 1 \\
                -1 & 0
            \end{pmatrix},& \mbox{for $\zeta\in\Gamma_3$.}
    \end{align}
    \item[(c)] $\Phi$ has the following behavior at
    infinity,
    \begin{equation}\label{RHP Phi: c}
        \Phi(\zeta)=(I+\bigO(1/\zeta))\zeta^{-\frac{1}{4}\sigma_3}N
        e^{-\theta(\zeta;x,T)\sigma_3},\qquad\mbox{as
        $\zeta\to\infty$,}
    \end{equation}
    where $N$ and $\theta$ are given by
    (\ref{definition: N theta}).
\end{itemize}

\begin{remark}
By multiplying $\Phi$ to the left with an appropriate matrix
independent of $\zeta$, see (\ref{definition: Q}) below, we obtain
by Proposition \ref{proposition: asymptotics Psihat} the RH problem
for $\Psi$, as stated in Section \ref{subsection: lax pair}, for the
particular choice of Stokes multipliers $s_1=s_2=s_5=s_6=0$,
$s_0=1$, and $s_3=s_4=-1$.
\end{remark}

\begin{remark}\label{remark: uniqueness Phi}
    Let $\Phi$ be a solution of the RH problem. By using the jump relations
    (\ref{RHP Phi: b1})--(\ref{RHP Phi: b3}) one has that $\det\Phi_+=\det\Phi_-$
    on $\Gamma$. This yields that $\det\Phi$ is entire. From (\ref{RHP Phi: c})
    we have that $\det\Phi(\zeta)\to 1$ as $\zeta\to\infty$, and thus, by
    Liouville's theorem, we have that $\det\Phi\equiv 1$.

    Now, suppose that $\tilde\Phi$ is a second solution of the RH
    problem. Then, since $\tilde\Phi$ and $\Phi$ satisfy the same
    jump relations on $\Gamma$, one has that $\tilde\Phi\,\Phi^{-1}$ is entire
    (observe that $\Phi^{-1}$ exists since $\det\Phi\equiv 1$). From
    (\ref{RHP Phi: c}) we have that $\tilde\Phi(\zeta)\Phi(\zeta)^{-1}\to
    I$ as $\zeta\to\infty$, and thus, by Liouville's theorem, we
    have that $\tilde\Phi\,\Phi^{-1}\equiv I$. We now have shown
    that if the RH problem for $\Phi$ has a solution, then this
    solution is unique.
\end{remark}

\subsection{Solvability of the RH problem for $\Phi$}
\label{subsection: solvability RH problem for Phi}

Here, our goal is to prove that the RH problem for $\Phi$ is
solvable for $x,T\in\mathbb R$. Moreover, we will also strengthen
the asymptotic condition (c) of the RH problem and prove analyticity
properties in the variables $x$ and $T$. In case $x=T=0$, the
solvability of the RH problem for $\Phi$ has been proven by Deift et
al.\ in \cite[Section 5.3]{DKMVZ2}. The general case is analogous
but for the convenience of the reader we will recall the different
steps in the proof and indicate where we need the restriction to
$x,T\in\mathbb{R}$. The result of this subsection is the following
lemma.

\begin{lemma}\label{lemma: solvability Phi}
    For every $x_0,T_0\in\mathbb{R}$, there exist neighborhoods $\mathcal V$
    of $x_0$ and $\mathcal W$ of $T_0$ such that for all $x\in\mathcal V$
    and $T\in\mathcal W$ the following holds.
    \begin{itemize}
        \item[(i)] The RH problem for $\Phi$ is solvable.
        \item[(ii)] The solution $\Phi$ of the RH problem for $\Phi$ has a full
            asymptotic expansion in powers of $\zeta^{-1}$ as follows,
            \begin{equation}\label{asymptotic expansion: Phi}
                \Phi(\zeta;x,T) \sim
                \Bigl(I+\sum_{k=1}^\infty A_k\zeta^{-k}\Bigr)
                \zeta^{-\frac{1}{4}\sigma_3} N e^{-\theta(\zeta;x,T)\sigma_3},
            \end{equation}
            as $\zeta\to\infty$, uniformly in $\mathbb{C}\setminus\Gamma$.
            Here, the $A_k=A_k(x,T)$ are real-valued for $x,T\in\mathbb{R}$.
        \item[(iii)] The solution $\Phi$ of the RH problem for $\Phi$, as well as
            the $A_k$ in {\rm(\ref{asymptotic expansion: Phi})}, are analytic both as
            functions of $x$ and $T$.
    \end{itemize}
\end{lemma}

\begin{remark}
    The important feature of this lemma is the following. In the next subsection we
    will show that $y=2A_{1,11}-A_{1,12}^2$, where $A_{1,ij}$ is the $(i,j)$-th entry
    of $A_1$, is a solution to the $P_I^2$ equation. From the above lemma
    we then have that this $y$ is real-valued and pole-free on the real
    axis, so that the first part of Theorem \ref{main theorem} is
    proven.
\end{remark}

In order to prove Lemma \ref{lemma: solvability Phi}, we transform,
as in \cite[Section 5.3]{DKMVZ2}, the RH problem for $\Phi$ into an
equivalent RH problem for $\widehat\Phi$ such that the jump matrix
for $\widehat\Phi$ is continuous on $\Gamma$ and converges
exponentially to the identity matrix as $\zeta\to\infty$ on
$\Gamma$, and such that the RH problem for $\widehat\Phi$ is
normalized at infinity. To do this, we introduce an auxiliary
$2\times 2$ matrix valued function $M$ satisfying the following RH
problem on a contour $\Gamma^\sigma=\bigcup_{j=1}^4\Gamma_j^\sigma$
consisting of four straight rays
\begin{equation}\label{contourgammasigma}
    \Gamma_1^\sigma:\arg\zeta=0, \qquad
    \Gamma_2^\sigma:\arg\zeta=\sigma, \qquad
    \Gamma_3^\sigma:\arg\zeta=\pi, \qquad
    \Gamma_4^\sigma:\arg\zeta=-\sigma,
\end{equation}
where $\sigma\in(\frac{\pi}{3},\pi)$. We orientate the straight
rays from the left to the right, as shown in Figure \ref{figure:
RHP Psi} for the contour $\Gamma$. The dependence on the parameter
$\sigma$ is needed in Section \ref{section: asymptotics y}. In
this section, we take $\sigma=6\pi/7$ fixed, so that
$\Gamma^\sigma=\Gamma$.

\subsubsection*{RH problem for $M$:}
\begin{itemize}
    \item[(a)] $M$ is analytic in $\mathbb{C}\setminus\Gamma^\sigma$.
    \item[(b)] $M$ satisfies the following jump relations on $\Gamma^\sigma$,
        \begin{align}
            \label{RHP M: b1}
            M_+(\zeta) &= M_-(\zeta)
                \begin{pmatrix}
                    1 & e^{-\frac{4}{3}\zeta^{3/2}}\\
                    0 & 1
                \end{pmatrix},
                &\mbox{for $\zeta\in\Gamma_1^\sigma$,} \\[1ex]
            \label{RHP M: b2}
            M_+(\zeta) &= M_-(\zeta)
                \begin{pmatrix}
                    1 & 0\\
                    e^{\frac{4}{3}\zeta^{3/2}} & 1
                \end{pmatrix},
                &\mbox{for $\zeta\in\Gamma_2^\sigma\cup\Gamma_4^\sigma$,} \\[1ex]
            \label{RHP M: b3}
            M_+(\zeta) &= M_-(\zeta)
                \begin{pmatrix}
                    0 & 1\\
                    -1 & 0
                \end{pmatrix},
                &\mbox{for $\zeta\in\Gamma_3^\sigma$.}
        \end{align}
    \item[(c)] $M$ has the following behavior at infinity,
        \begin{equation}\label{RHP M: c}
            M(\zeta)\sim\Bigl(I+\sum_{k=1}^\infty B_k\zeta^{-k}\Bigr)
                \zeta^{-\frac{1}{4}\sigma_3}N, \qquad
                \mbox{as $\zeta\to\infty$,}
        \end{equation}
        uniformly for $\zeta\in\mathbb{C}\setminus\Gamma^\sigma$ and $\sigma$ in compact subsets
        of $(\frac{\pi}{3},\pi)$. Here, $N$ is given by equation (\ref{definition: N theta}),
        and for $k\geq 1$,
        \begin{equation}\label{definition: Bk}
            B_{3k-2}=
                \begin{pmatrix}
                    0 & 0 \\
                    t_{2k-1} & 0
                \end{pmatrix}, \qquad
            B_{3k-1}=
                \begin{pmatrix}
                    0 & s_{2k-1} \\
                    0 & 0
                \end{pmatrix}, \qquad
            B_{3k}=
                \begin{pmatrix}
                    s_{2k} & 0 \\
                    0 & t_{2k}
                \end{pmatrix},
        \end{equation}
        with
        \begin{equation}\label{definition: sk tk}
            s_k = \frac{\Gamma(3k+1/2)}{36^k k! \Gamma(k+1/2)}, \qquad
            t_k = -\frac{6k+1}{6k-1}s_k.
        \end{equation}
\end{itemize}
It is well-known, see e.g.~\cite{Deift,DKMVZ1}, that there exists a
unique solution $M$ to the above RH problem given in terms of Airy
functions $\Ai$. The matrix valued function $M$ is the so-called
Airy parametrix and for the purpose of this paper we will not need
its exact expression but refer the reader to \cite{Deift,DKMVZ1} for
this.

We now define $\widehat\Phi(\zeta;x,T)=\widehat\Phi(\zeta)$ by
\begin{equation}\label{Psihat in Psi}
    \widehat\Phi(\zeta)=\Phi(\zeta)e^{\theta(\zeta)\sigma_3}
    M(\zeta)^{-1},\qquad\mbox{for
    $\zeta\in\mathbb{C}\setminus\Gamma$.}
\end{equation}
A straightforward calculation, using (\ref{RHP Phi: b1})--(\ref{RHP
Phi: c}), (\ref{RHP M: b1})--(\ref{RHP M: c}), and
$\theta_+(\zeta)+\theta_-(\zeta)=0$ for $\zeta\in\mathbb{R}_-$,
shows that $\widehat\Phi$ satisfies the following RH problem.

\subsubsection*{RH problem for $\widehat\Phi$:}

\begin{itemize}
    \item[(a)] $\widehat\Phi$ is analytic in
    $\mathbb{C}\setminus\Gamma$.
    \item[(b)] $\widehat\Phi_+(\zeta)=\widehat\Phi_-(\zeta)\hat v(\zeta)$
    for $\zeta\in\Gamma$, where $v(\zeta)=v(\zeta;x,T)$ is
    given by
    \begin{equation}\label{RHP Psihat: b}
        v(\zeta)=
        \begin{cases}
            M_-(\zeta)
            \begin{pmatrix}
                1 &
                e^{-2\theta(\zeta)}-e^{-\frac{4}{3}\zeta^{3/2}}\\
                0 & 1
            \end{pmatrix} M_-(\zeta)^{-1},& \mbox{for
            $\zeta\in\Gamma_1$,}\\[4ex]
            M_-(\zeta)
            \begin{pmatrix}
                1 & 0 \\
                e^{2\theta(\zeta)}-e^{\frac{4}{3}\zeta^{3/2}} & 1
            \end{pmatrix} M_-(\zeta)^{-1},& \mbox{for
            $\zeta\in\Gamma_2\cup\Gamma_4$,}\\[4ex]
            I, & \mbox{for $\zeta\in\Gamma_3$.}
        \end{cases}
    \end{equation}
    \item[(c)] $\widehat\Phi(\zeta)=I+\bigO(1/\zeta)$,\qquad as
        $\zeta\to\infty$.
\end{itemize}
Observe that the jump matrix $v$ is indeed continuous on $\Gamma$
and that it converges exponentially to the identity matrix as
$\zeta\to\infty$ on $\Gamma$. This RH problem corresponds to the RH
problem \cite[(5.108)--(5.110)]{DKMVZ2}, and the only difference is
that we now have a factor $e^{\pm 2\theta}$ (containing the $x,T$
dependence) instead of $e^{\pm\zeta^{(4\nu+3)/2}}$ in the jump
matrices.

\begin{varproof}\textbf{of Lemma \ref{lemma: solvability Phi} (i).}
From (\ref{Psihat in Psi}) it follows that proving the solvability
of the RH problem for $\Phi$ is equivalent to proving the
solvability of the RH problem for $\widehat\Phi$. By general theory
of the construction of solutions of RH problems, this is reduced to
the study of the singular integral operator,
\begin{equation}
    C_v:L^2(\Gamma)\to L^2(\Gamma):f\mapsto C_+\left[f\left(I-v^{-1}\right)\right],
\end{equation}
where $v$ is the jump matrix (\ref{RHP Psihat: b}) of the RH problem
for $\widehat\Phi$, and where $C_+$ is the $+$boundary value of the
Cauchy operator
\[
    Cf(z)=\frac{1}{2\pi
    i}\int_\Gamma\frac{f(s)}{s-z}ds,\qquad\mbox{for $z\in\mathbb{C}\setminus\Gamma$.}
\]
Indeed, suppose that $I-C_v$ is invertible in $L^2(\Gamma)$. Then,
there exists $\mu\in L^2(\Gamma)$ such that
$(I-C_v)\mu=C_+(I-v^{-1})$, and it is immediate that
\begin{equation}\label{solution: RHP Psihat}
    \widehat\Phi(\zeta)\equiv I+\frac{1}{2\pi i}\int_\Gamma\frac{(I+\mu(s))(I-
    v(s)^{-1})}{s-\zeta}ds,\qquad\mbox{for
    $\zeta\in\mathbb{C}\setminus\Gamma$,}
\end{equation}
is analytic in $\mathbb{C}\setminus\Gamma$ and satisfies (since $C_+
- C_-=I$) condition (b) of the RH problem for $\widehat\Phi$ in the
so-called $L^2$-sense. However, as in \cite[Step 3 of Sections 5.2
and 5.3]{DKMVZ2}, one can use the analyticity of $v$ to show that
$\widehat\Phi$ satisfies jump condition (b) in the sense of
continuous boundary values, as well. Further, as in
\cite[Proposition 5.4]{DKMVZ2}, it follows from the exponential
decaying of $I-v^{-1}$ as $\zeta\to\infty$ on $\Gamma$ that the
asymptotic condition (c) of the RH problem for $\widehat\Phi$ is
also satisfied. We summarize that the RH problem for $\widehat\Phi$
is solvable, with solution given by (\ref{solution: RHP Psihat}),
provided the singular integral operator $I-C_v$ is invertible in
$L^2(\Gamma)$.

\smallskip

First, we consider the case $x,T\in\mathbb{R}$. For this case, we
show that $I-C_v$ is invertible by showing that it is a Fredholm
operator with zero index and kernel $\{0\}$. Exactly as in
\cite[Steps 1 and 2 of Section 5.3]{DKMVZ2} one has that $I-C_v$
is a Fredholm operator with zero index. In this step, one does not
need the restriction to real $x$ and $T$. It remains to prove that
the kernel of $I-C_v$ is $\{0\}$, and it is in this step that we
will need the restriction that $x,T\in\mathbb R$. This is (again)
as in \cite[Section 5.3]{DKMVZ2} but for the convenience of the
reader we will indicate were we need $x$ and $T$ to be real.

Suppose there exists $\mu_0\in L^2(\Gamma)$ such that
$(I-C_v)\mu_0=0$. One can then show that the matrix valued function
$\widehat \Phi_0$ defined by
\begin{equation}\label{definition: Psi0hat}
    \widehat\Phi_0(\zeta)\equiv\frac{1}{2\pi i}
    \int_\Gamma\frac{\mu_0(s)(I-v(s)^{-1})}{s-\zeta}ds,\qquad\mbox{for
    $\zeta\in\mathbb{C}\setminus\Gamma$,}
\end{equation}
is a solution to the RH problem for $\widehat\Phi$, but with the
asymptotic condition (c) replaced by the \textit{homogeneous}
condition
\begin{equation}\label{RHP Psi0hat: c}
    \widehat\Phi_0(\zeta)=\bigO(1/\zeta),\qquad\mbox{as
    $\zeta\to\infty$, uniformly for
    $\zeta\in\mathbb{C}\setminus\Gamma$.}
\end{equation}
Since $\mu_0=\widehat\Phi_{0,+}$ (which follows from
(\ref{definition: Psi0hat}) together with $(I-C_v)\mu_0=0$), we
need to show that $\widehat\Phi_0\equiv 0$. Showing that a
solution of the homogeneous RH problem is identically zero, is
known in the literature as a vanishing lemma, see
\cite{DKMVZ2,FokasMuganZhou,FokasZhou}.

Now, let
\[
    \Phi_0(\zeta)=\widehat\Phi_0(\zeta)
    M(\zeta),\qquad\mbox{for $\zeta\in\mathbb{C}\setminus\Gamma$,}
\]
then it is straightforward to check, using (\ref{RHP M:
b1})--(\ref{RHP M: c}), (\ref{RHP Psihat: b}), and (\ref{RHP
Psi0hat: c}), that $\Phi_0$ solves the following RH problem.

\subsubsection*{RH problem for $\Phi_0$:}

\begin{itemize}
    \item[(a)] $\Phi_0$ is analytic in $\mathbb{C}\setminus\Gamma$.
    \item[(b)] $\Phi_0$ satisfies the following jump relations on $\Gamma$,
    \begin{align}
        \label{RHP Psi0: b1}
        \Phi_{0,+}(\zeta)&=\Phi_{0,-}(\zeta)\begin{pmatrix}
                1 & e^{-2\theta(\zeta)}\\
                0 & 1
            \end{pmatrix},& \mbox{for $\zeta\in\Gamma_1$,}\\[1ex]
        \label{RHP Psi0: b2}
        \Phi_{0,+}(\zeta)&=\Phi_{0,-}(\zeta)\begin{pmatrix}
                1 & 0 \\
                e^{2\theta(\zeta)} & 1
            \end{pmatrix},&\mbox{for
            $\zeta\in\Gamma_2\cup\Gamma_4$,}\\[1ex]
        \label{RHP Psi0: b3}
        \Phi_{0,+}(\zeta)&=\Phi_{0,-}(\zeta)\begin{pmatrix}
                0 & 1 \\
                -1 & 0
            \end{pmatrix},& \mbox{for $\zeta\in\Gamma_3$.}
    \end{align}
    \item[(c)] $\Phi_0(\zeta)=\bigO(1/\zeta)\zeta^{-\frac{1}{4}\sigma_3}N$,
    \qquad as $\zeta\to\infty$, uniformly for
    $\zeta\in\mathbb{C}\setminus\Gamma$.
\end{itemize}

Further, we introduce an auxiliary matrix valued function $A$ with
jumps only on $\mathbb{R}$, as follows, cf.~\cite[Equations
(5.135)--(5.138)]{DKMVZ2}
\begin{equation}
    A(\zeta)=
    \begin{cases}
        \Phi_0(\zeta)
        \begin{pmatrix}
            0 & -1 \\
            1 & 0
        \end{pmatrix},& \mbox{for $0<\arg\zeta<\frac{6\pi}{7}$,}
        \\[4ex]
        \Phi_0(\zeta)
        \begin{pmatrix}
            1 & 0 \\
            e^{2\theta(\zeta)} & 1
        \end{pmatrix}
        \begin{pmatrix}
            0 & -1 \\
            1 & 0
        \end{pmatrix},& \mbox{for
        $\frac{6\pi}{7}<\arg\zeta<\pi$,}\\[4ex]
        \Phi_0(\zeta)
        \begin{pmatrix}
            1 & 0 \\
            -e^{2\theta(\zeta)} & 1
        \end{pmatrix},& \mbox{for
        $-\pi<\arg\zeta<-\frac{6\pi}{7}$,}\\[4ex]
        \Phi_0(\zeta), & \mbox{for $-\frac{6\pi}{7}<\arg\zeta<0$.}
    \end{cases}
\end{equation}
Using (\ref{RHP Psi0: b1})--(\ref{RHP Psi0: b3}) and condition (c)
of the RH problem for $\Phi_0$ one can then check that $A$ is a
solution to the following RH problem.

\subsubsection*{RH problem for $A$:}

\begin{itemize}
    \item[(a)] $A$ is analytic in $\mathbb C\setminus\mathbb R$
    \item[(b)] $A$ satisfies
    the following jump relation on $\mathbb{R}$,
    \begin{align}
        \label{RHP A: a1}
        A_+(\zeta)&=A_-(\zeta)\begin{pmatrix}
            1 & -e^{2\theta_+(\zeta)} \\
            e^{2\theta_-(\zeta)} & 0
        \end{pmatrix},&\mbox{for $\zeta\in\mathbb R_-$,}\\[1ex]
        \label{RHP A: a2}
        A_+(\zeta)&=A_-(\zeta)\begin{pmatrix}
            e^{-2\theta(\zeta)} & -1 \\
            1 & 0
        \end{pmatrix},&\mbox{for $\zeta\in\mathbb R_+$,}
    \end{align}
    \item[(c)] $A(\zeta)=\bigO(\zeta^{-3/4})$,\qquad as
    $\zeta\to\infty$, uniformly for $\zeta\in\mathbb C\setminus\mathbb
    R$.
\end{itemize}

Now, we define $Q(\zeta)=A(\zeta)A^*(\bar\zeta)$, where $A^*$
denotes the Hermitian conjugate of $A$. The matrix valued function
$Q$ is analytic in the upper half plane, continuous up to
$\mathbb{R}$, and decays like $\zeta^{-3/2}$ as $\zeta\to\infty$. By
Cauchy's theorem this implies, $\int_{\mathbb{R}}Q_+(s)ds=0$. Using
the jump relations (\ref{RHP A: a1}) and (\ref{RHP A: a2}) we then
have,
\[
    \int_{\mathbb{R}_-}A_-(s)
    \begin{pmatrix}
        1 & -e^{2\theta_+(s)} \\
        e^{2\theta_-(s)} & 0
    \end{pmatrix}A^*_-(s)ds+\int_{\mathbb{R}_+}A_-(s)
        \begin{pmatrix}
            e^{-2\theta(s)} & -1 \\
            1 & 0
        \end{pmatrix}
        A^*_-(s)ds=0.
\]
Adding this to its Hermitian conjugate, and using the fact
$\overline{\theta_+(s)}=\theta_-(s)$ for $s\in\mathbb{R}_-$ (which
is true since $x,T\in\mathbb{R}$), we arrive at, cf.~\cite[Equation
(5.146)]{DKMVZ2}
\begin{equation}
    \int_{\mathbb{R}_-}A_-(s)
    \begin{pmatrix}
        2 & 0 \\
        0 & 0
    \end{pmatrix}A^*_-(s)ds+\int_{\mathbb{R}_+}A_-(s)
        \begin{pmatrix}
            2e^{-2\theta(s)} & -1 \\
            1 & 0
        \end{pmatrix}
        A^*_-(s)ds=0.
\end{equation}
This is the crucial step where we need $x$ and $T$ to be real. The latter
relation implies that the first column of $A_-$ is identically zero, and the
jump relations (\ref{RHP A: a1}) and (\ref{RHP A: a2}) then imply that the
second column of $A_+$ is identically zero, as well.

By writing out the RH conditions for each entry of $A$ and using
the vanishing of the first column of $A_-$ and the second column
of $A_+$, the matrix RH problem reduces to two scalar RH problems.
The proof that the solutions of those scalar RH problems (and thus
also the second column of $A_-$ and the first column of $A_+$) are
identically zero, is exactly as in \cite[Step 3 of Section
5.3]{DKMVZ2} using Carlson's theorem, see \cite{ReedSimon}, and we
will not go into detail about this. We then have shown that
$A\equiv 0$, so that also $\widehat\Phi_0\equiv 0$ and thus
$\mu_0\equiv 0$. We now have proven that $I-C_v$ is invertible for
$x,T\in\mathbb{R}$, which implies that the RH problem for
$\widehat\Phi$ (and thus also the RH problem for $\Phi$) is
solvable for $x,T\in\mathbb{R}$.

\medskip

Next, fix $x_0,T_0\in\mathbb{R}$. Above, we have shown that the
singular integral operator $I-C_{v(\cdot\, ;\, x_0,T_0)}$ is
invertible. Since
\[
    I-C_{v(\cdot\, ;\, x,T)}=\left(I-C_{v(\cdot\, ;\, x_0,T_0)}\right)
    \left[I+\left(I-C_{v(\cdot\, ;\, x_0,T_0)}\right)^{-1}\left(C_{v(\cdot\, ;\, x_0,T_0)}-C_{
    v(\cdot\, ;\, x,T)}\right)\right],
\]
it then follows that $I-C_{v(\cdot\, ;\, x,T)}$ is invertible
provided
\[
    \bigl\|\left(I-C_{v(\cdot\, ;\, x_0,T_0)}\right)^{-1}\left(C_{v(\cdot\, ;\, x_0,T_0)}-C_{
    v(\cdot\, ;\, x,T)}\right)\bigr\| < 1,.
\]
where $\|\cdot\|$ denotes the operator norm. It is straightforward
to check that there exist neighborhoods $\mathcal V$ of $x_0$ and
$\mathcal W$ of $T_0$ such that for all $x\in\mathcal V$ and
$T\in\mathcal W$,
\begin{align*}
    \bigl\|C_{v(\cdot\, ;\, x_0,T_0)}-C_{v(\cdot\, ;\, x,T)}\bigr\|
    &\leq \bigl\|C_+ \bigr\|\  \bigl\|v(\cdot\, ;\, x_0,T_0)-
    v(\cdot\, ;\, x,T)\bigr\|_{L^{(\infty)}(\Gamma)}\\[1ex]
    &<\bigl\|\left(I-C_{
    v(\cdot\, ;\, x_0,T_0)}\right)^{-1}\bigr\|^{-1},
\end{align*}
which implies that the operator $I-C_{v(\cdot\, ;\,x,T)}$ is
invertible. Hence the RH problem for $\widehat\Phi$, and thus also
the RH problem for $\Phi$, is solvable for $x\in\mathcal V$ and
$T\in\mathcal W$. This finishes the proof of the first part of the
lemma.
\end{varproof}

\begin{varproof}\textbf{of Lemma \ref{lemma: solvability Phi} (ii).}
It follows from the asymptotic expansion (\ref{RHP M: c}) of $M$
together with $\Phi=\widehat\Phi M e^{-\theta\sigma_3}$, see
(\ref{Psihat in Psi}), that we need to show that $\widehat\Phi$ has
a full asymptotic expansion in powers of $\zeta^{-1}$. Insert the
relation
\[
    \frac{1}{s-\zeta}=-\sum_{k=1}^n
s^{k-1}\zeta^{-k}+\frac{s^n}{\zeta^n(s-\zeta)},\qquad\mbox{for
$n\in\mathbb{N}$,}
\]
into the solution $\widehat\Phi$ of the RH problem for
$\widehat\Phi$, which is given by (\ref{solution: RHP Psihat}). We
then obtain for any $n\in\mathbb{N}$,
\begin{equation}\label{Psihat: expanded}
    \widehat\Phi=I+\sum_{k=1}^n \widehat B_k\zeta^{-k}+\frac{1}{2\pi i}\int_\Gamma \frac{s^n(I+\mu(s))(I-
    v(\zeta)^{-1})}{\zeta^n(s-\zeta)}ds,
\end{equation}
where
\begin{equation}\label{definition: Bkhat}
    \widehat B_k=-\frac{1}{2\pi i}\int_\Gamma s^{k-1}(I+\mu(s))(I-
    v(s)^{-1})ds.
\end{equation}
As in \cite[Proposition 5.4]{DKMVZ2} one can check that the integral
in (\ref{Psihat: expanded}) is of order $\bigO(\zeta^{-(n+1)})$ as
$\zeta\to\infty$ uniformly for $\zeta\in\mathbb{C}\setminus\Gamma$.
We then have shown that $\widehat\Phi$ has the following asymptotic
expansion in powers of $\zeta^{-1}$,
\begin{equation}\label{asymptotic expansion: Psihat}
    \widehat\Phi(\zeta)\sim I+\sum_{k=1}^\infty\widehat B_k\zeta^{-k},\qquad\mbox{as $\zeta\to\infty$, uniformly for
    $\zeta\in\mathbb{C}\setminus\Gamma$.}
\end{equation}
From (\ref{RHP M: c}), (\ref{asymptotic expansion: Psihat}), and the
fact that $\Phi=\widehat\Phi P e^{-\theta\sigma_3}$ it now follows
that $\Phi$ has a full asymptotic expansion in the form
(\ref{asymptotic expansion: Phi}), where (with $B_0=\widehat B_0=I$)
\begin{equation}\label{definition: Ak}
    A_k=\sum_{j=0}^k B_j\widehat B_{k-j}.
\end{equation}

It remains to show that the $A_k$ are real-valued for
$x,T\in\mathbb{R}$. It is straightforward to verify that for
$x,T\in\mathbb{R}$ the matrix valued function
$-i\overline{\Phi(\overline{\zeta};x,T)}\sigma_3$ is a solution to
the RH problem for $\Phi$. By uniqueness we then have
\[
    \Phi(\zeta;x,T)=-i\overline{\Phi(\overline{\zeta};x,T)}\sigma_3,\qquad\mbox{for
    $x,T\in\mathbb{R}$,}
\]
which yields
\[
    \Bigl(I+\sum_{k=1}^\infty A_k\zeta^{-k}\Bigr)\zeta^{-\frac{1}{4}\sigma_3}N
        e^{-\theta(\zeta;x,T)\sigma_3}=\Bigl(I+\sum_{k=1}^\infty \overline{A_k}\zeta^{-k}\Bigr)
        \zeta^{-\frac{1}{4}\sigma_3}N
        e^{-\theta(\zeta;x,T)\sigma_3},
\]
and hence $A_k=\overline{A_k}$ for $x,T\in\mathbb{R}$. This proves
the second part of the lemma.
\end{varproof}

\medskip

\begin{varproof}\textbf{of Lemma \ref{lemma: solvability Phi} (iii).}
    We show that $\Phi$ and $A_k$ are analytic in $x$, for $x\in\mathcal V$.
    The analyticity in $T$ follows in a similar fashion. In order to show that $\widehat\Phi$ (and thus also
    $\Phi$) is analytic for $x\in\mathcal V$ we
    need to show that, letting $h\to 0$ in the complex plane,
    \[
        \lim_{h\to 0}\frac{1}{h}\bigl(\widehat\Phi(\zeta;x+h,T)-\widehat\Phi(\zeta;x,T)\bigr)
    \]
    exists. Consider the $2\times 2$ auxiliary matrix valued
    function $H(\zeta;x,T;h)=H(\zeta)$ defined as follows,
    \begin{equation}
        H(\zeta)=\widehat\Phi(\zeta;x+h,T)\widehat\Phi(\zeta;x,T)^{-1},\qquad\mbox{for
        $\zeta\in\mathbb{C}\setminus\Gamma$.}
    \end{equation}
    Here we take $h$ sufficiently small, so that $\Phi(\zeta;x+h,T)$
    exists by part (i) of the lemma. It is straightforward to check that
    $H$ satisfies the following RH problem.

    \subsubsection*{RH problem for $H$:}
    \begin{itemize}
        \item[(a)] $H$ is analytic in $\mathbb{C}\setminus\Gamma$.
        \item[(b)] $H$ satisfies the jump relation
            $H_+(\zeta)=H_-(\zeta)v_H(\zeta)$ for $\zeta\in\Gamma$,
            where
            \begin{align*}
                v_H(\zeta)&=
                I+e^{-2\theta(\zeta;x,T)}(e^{-2h\zeta^{1/2}}-1)\widehat\Phi_-(\zeta;x,T)
            \begin{pmatrix} 0 & 1 \\ 0 & 0
            \end{pmatrix}\widehat\Phi_-(\zeta;x,T)^{-1},&
            \zeta\in\Gamma_1,
            \\[1ex]
            v_H(\zeta)&=
I+e^{2\theta(\zeta;x,T)}(e^{2h\zeta^{1/2}}-1)\widehat\Phi_-(\zeta;x,T)
            \begin{pmatrix} 0 & 0 \\ 1 & 0
            \end{pmatrix}\widehat\Phi_-(\zeta;x,T)^{-1},&
            \zeta\in\Gamma_2\cup\Gamma_4,
            \\[1ex]
            v_H(\zeta)&=I,&\zeta\in\Gamma_3.
            \end{align*}
    \item[(c)] $H(\zeta)=I+\bigO(1/\zeta)$,\qquad as $\zeta\to\infty$,
    uniformly for $\zeta\in\mathbb{C}\setminus\Gamma$.
\end{itemize}
Since $v_H(\zeta)=I+\bigO(h)$ as $h\to 0$ uniformly for
$\zeta\in\Gamma$, where the $\bigO(h)$-term can be expanded into a
full asymptotic expansion in powers of $h$, it follows as in
\cite{Deift,DKMVZ1,DKMVZ2} that
\begin{equation}
    \widehat\Phi(\zeta;x+h,T)\widehat\Phi(\zeta;x,T)^{-1}=
    H(\zeta)=I+hH_1(\zeta;x,T)+\bigO(h^2),\qquad \mbox{as $h\to 0$,}
\end{equation}
where $H_1$ is a $2\times 2$ matrix valued function independent of
$h$. This yields,
\[
    \lim_{h\to
0}\frac{1}{h}\bigl(\widehat\Phi(\zeta;x+h,T)-\widehat\Phi(\zeta;x,T)\bigr)=H_1(\zeta;x,T)\widehat\Phi(\zeta;x,T),
\]
which implies that $\widehat\Phi$ (and thus also $\Phi$) is analytic
for $x\in\mathcal V$.

It remains to show that the matrix valued functions $A_k$ are
analytic for $x\in\mathcal V$. By (\ref{solution: RHP Psihat}), it
is immediate that
\[
    \widehat\Phi_+(\zeta)=I+\mu(\zeta),\qquad \mbox{for
    $\zeta\in\Sigma$,}
\]
so that $\mu$ is analytic for $x\in\mathcal V$. By (\ref{definition:
Bkhat}) it then follows that $\widehat B_k$ is also analytic for
$x\in\mathcal V$. This yields, by (\ref{definition: Ak}) and
(\ref{definition: Bk}), the analyticity of $A_k$, and hence the last
part of the lemma is proven.
\end{varproof}

\subsection{Proof of Theorem \ref{main theorem} (i)}
\label{subsection: 2.3}

In order to prove the existence part of Theorem \ref{main theorem}
we proceed as follows. Introduce, for $x,T\in\mathbb{R}$, a
$2\times 2$ matrix valued function $\Psi(\zeta;x,T)=\Psi(\zeta)$
by multiplying the solution $\Phi$ of the RH problem for $\Phi$ to
the left with an appropriate matrix independent of $\zeta$,
\begin{equation}\label{definition: Q}
    \Psi(\zeta)=\begin{pmatrix}
        1 & 0 \\
        A_{1,12} & 1
    \end{pmatrix}\Phi(\zeta),\qquad\mbox{for $\zeta\in\mathbb{C}\setminus\Gamma$.}
\end{equation}
Here $A_{1,12}$ is the $(1,2)$-entry of the $2\times 2$ matrix
$A_1=A_1(x,T)$ appearing in the asymptotic expansion
(\ref{asymptotic expansion: Phi}) of $\Phi$ at infinity. The
important feature of this transformation is that $\Psi$ satisfies
the RH problem for $P_I^2$, see Section \ref{subsection: lax
pair}, as we will show in the following proposition.

\begin{proposition}\label{proposition: asymptotics Psihat}
    The matrix valued function $\Psi$, defined by {\rm (\ref{definition:
    Q})}, is a solution to the RH problem for $\Psi$, see Section
    {\rm \ref{subsection: lax pair}}, with Stokes
    multipliers $s_1=s_2=s_5=s_6=0$, $s_0=1$, and $s_3=s_4=-1$,
    and with the asymptotic condition (c) replaced by the stronger
    condition
    \begin{equation}\label{definition: Qhat}
        \Psi(\zeta) = \zeta^{-\frac{1}{4}\sigma_3}N\widehat \Psi(\zeta)e^{-\theta(\zeta)\sigma_3},
    \end{equation}
    where $\widehat \Psi$ has a full asymptotic expansion in powers of $\zeta^{-1/2}$ as follows,
    \begin{multline}\label{asymptotics Qhat}
        \widehat \Psi(\zeta;x,T)\sim
        I-h\sigma_3\zeta^{-1/2}+\frac{1}{2}\begin{pmatrix}h^2 & iy\\-iy &
        h^2\end{pmatrix}\zeta^{-1}\\[1ex]+\frac{1}{2}\sum_{k=1}^\infty\left[
        \begin{pmatrix}q_k & i r_k \\ ir_k &
        -q_k\end{pmatrix}\zeta^{-k-\frac{1}{2}}+\begin{pmatrix}v_k & iw_k \\
        -iw_k & v_k\end{pmatrix}\zeta^{-k-1}\right],
    \end{multline}
    as $\zeta\to\infty$ uniformly for
    $\zeta\in\mathbb{C}\setminus\Gamma$. Here, $y=y(x,T)$ is given by
    \begin{equation}\label{definition: y}
        y=2A_{1,11}-A_{1,12}^2.
    \end{equation}
    Further, $h=A_{1,12}$ and the $q_k, r_k, v_k$
    and $w_k$ are some unimportant functions of $x$ and $T$ (independent of $\zeta$).
\end{proposition}

\begin{proof}
    The fact that $\Psi$ satisfies conditions (a) and (b) of the RH
    problem for $\Psi$ follows trivially from (\ref{definition: Q}) together with conditions (a) and (b)
    of the RH problem for $\Phi$. So, it remains to show that
    $\widehat\Psi$ given by
    \begin{equation}\label{proof: proposition asymptotics Psihat: eq1a}
        \widehat\Psi=N^{-1}\zeta^{\frac{\sigma_3}{4}}\Psi(\zeta)e^{\theta(\zeta)\sigma_3},
    \end{equation}
    satisfies an asymptotic expansion of the form (\ref{asymptotics Qhat}) with $y$ given
    by (\ref{definition: y}). It follows from (\ref{proof: proposition asymptotics Psihat:
    eq1a}), (\ref{definition: Q}) and (\ref{asymptotic expansion: Phi}) that
    \begin{align}\label{proof: proposition asymptotics Psihat: eq1}
        \nonumber
        \widehat\Psi(\zeta) &\sim
        N^{-1}\zeta^{\frac{\sigma_3}{4}}
        \begin{pmatrix}
            1 & 0 \\
            A_{1,12} & 1
        \end{pmatrix}
        \left[I+\sum_{k=1}^\infty
        A_k\zeta^{-k}\right]\zeta^{-\frac{\sigma_3}{4}}N \\[1ex]
        &\sim
        N^{-1}\left(\sum_{k=0}^\infty \zeta^{\frac{\sigma_3}{4}}\tilde A_k
        \zeta^{-\frac{\sigma_3}{4}}\zeta^{-k}\right)N,
    \end{align}
    where
    \[
        \tilde A_0=\begin{pmatrix}
            1 & 0 \\
            A_{1,12} & 1
        \end{pmatrix},\qquad\mbox{and}\quad
        \tilde A_k=\begin{pmatrix}
            1 & 0 \\
            A_{1,12} & 1
        \end{pmatrix}A_k,\qquad\mbox{for $k\geq 1$.}
    \]
    Now, using the facts that $\tilde A_{0,11}=\tilde A_{0,22}=1$,
    that $\tilde A_{0,12}=0$ and that $\tilde A_{0,21}=\tilde
    A_{1,12}=A_{1,12}$ we find,
        \begin{align*}
        & \sum_{k=0}^\infty
        \zeta^{\frac{\sigma_3}{4}}
        \tilde A_k\zeta^{-\frac{\sigma_3}{4}}\zeta^{-k}\\
        &\qquad\qquad=\sum_{k=0}^\infty\left[
        \begin{pmatrix}
            0 & 0 \\
            \tilde A_{k,21} & 0
        \end{pmatrix}\zeta^{-k-\frac{1}{2}}+
        \begin{pmatrix}
            0 & \tilde A_{k,12}\\ 0 & 0
        \end{pmatrix}\zeta^{-k+\frac{1}{2}}
        +\begin{pmatrix}
            \tilde A_{k,11} & 0 \\ 0 & \tilde A_{k,22}
        \end{pmatrix}\zeta^{-k}\right]\\[2ex]
        &\qquad\qquad=I+A_{1,12}
        \begin{pmatrix}
            0 & 1 \\
            1 & 0
        \end{pmatrix}\zeta^{-1/2}+\begin{pmatrix}
            \tilde A_{1,11} & 0 \\
            0 & \tilde A_{1,22}
        \end{pmatrix}\zeta^{-1}\\[1ex]
        &\qquad\qquad\qquad\qquad+\sum_{k=1}^\infty\left[
        \begin{pmatrix}
            0 & \tilde A_{k+1,12} \\
            \tilde A_{k,21} & 0
        \end{pmatrix}\zeta^{-k-\frac{1}{2}}+
        \begin{pmatrix}
            \tilde A_{k+1,11} & 0 \\
            0 & \tilde A_{k+1,22}
        \end{pmatrix}\zeta^{-k-1}\right].
    \end{align*}
    Inserting this into (\ref{proof: proposition asymptotics Psihat: eq1}) and using (\ref{definition: N theta})
    we arrive at,
    \begin{multline}\label{proof: proposition asymptotics Psihat: eq2}
        \widehat \Psi(\zeta;x,T)\sim
        I-h\sigma_3\zeta^{-1/2}+\frac{1}{2}\begin{pmatrix}\tilde A_{1,11}+\tilde A_{1,22} &
        i(\tilde A_{1,11}-\tilde A_{1,22})\\-i(\tilde A_{1,11}-\tilde A_{1,22}) &
        \tilde A_{1,11}+\tilde A_{1,22}\end{pmatrix}\zeta^{-1}\\[1ex]+\frac{1}{2}\sum_{k=1}^\infty\left[
        \begin{pmatrix}q_k & i r_k \\ ir_k &
        -q_k\end{pmatrix}\zeta^{-k-\frac{1}{2}}+\begin{pmatrix}v_k & iw_k \\
        -iw_k & v_k\end{pmatrix}\zeta^{-k-1}\right],
    \end{multline}
    where $h=A_{1,12}$ and where the $q_k, r_k, v_k$, and $w_k$ can be
    written down explicitly in terms of $\tilde A_k$ and $\tilde
    A_{k+1}$. Now, note that since $\det\Phi\equiv 1$ (see Remark \ref{remark: uniqueness Phi})
    and since, by (\ref{asymptotic expansion: Phi}),
    \[
        \det\Phi=1+(A_{1,11}+A_{1,22})\zeta^{-1}+\bigO(\zeta^{-2}),\qquad\mbox{as
        $\zeta\to\infty$,}
    \]
    we have that $A_{1,22}=-A_{1,11}$. This together with the facts that $\tilde A_{1,11}=A_{1,11}$
    and $\tilde A_{1,22}=A_{1,12}^2+A_{1,22}$ yields,
    \[
        \tilde A_{1,11}+\tilde A_{1,22}=A_{1,12}^2=h^2,\qquad \tilde A_{1,11}-\tilde
        A_{1,22}=2A_{1,11}-A_{1,12}^2=y.
    \]
    Inserting this into (\ref{proof: proposition asymptotics Psihat: eq2}) the proposition is proven.
\end{proof}

The idea is now to show that $\Psi$ satisfies the linear system of
differential equations (\ref{differential
equations})--(\ref{introduction: W}) with $y$ given by
(\ref{definition: y}), so that by compatibility of the Lax pair
this $y$ is a solution to the $P_I^2$ equation (\ref{PI2}). Since
by Lemma \ref{lemma: solvability Phi} the functions $A_{1,11}$ and
$A_{1,12}$ are real-valued and pole-free for $x,T\in\mathbb{R}$ we
have that $y$ itself is real-valued and pole-free for
$x,T\in\mathbb{R}$, so that the first part of Theorem \ref{main
theorem} is proven.

\begin{varproof}\textbf{of Theorem \ref{main theorem} (i).}
    Recall from the above discussion that we need to show that
    the matrix valued functions  (note that, by
    Lemma \ref{lemma: solvability Phi} (iii) and (\ref{definition: Q}),
    $\Psi$ is differentiable with respect to $x$)
    \begin{equation}\label{definition: U and W}
        U=\frac{\partial \Psi}{\partial \zeta}\,
            \Psi^{-1}\qquad\mbox{and}\qquad
        W=\frac{\partial \Psi}{\partial x}\, \Psi^{-1},
    \end{equation}
    are of the form (\ref{introduction: U}) and
    (\ref{introduction: W}), respectively, with $y$ given by (\ref{definition: y}).
    Observe that, since $\Psi$ has constant jump matrices, the derivatives
    $\frac{\partial \Psi}{\partial \zeta}$ and $\frac{\partial \Psi}{\partial x}$ have the same
    jumps as $\Psi$, and hence $U$ and $W$ are entire.

    First, we focus on $U$. By (\ref{definition: Qhat}),
    \begin{equation}\label{proof: main theorem: U eq1}
        U=-\frac{\partial\theta}{\partial\zeta}\, \zeta^{-\frac{\sigma_3}{4}}\left(N\widehat
        \Psi\sigma_3\widehat \Psi^{-1}N^{-1}\right)\zeta^{\frac{\sigma_3}{4}}+
        \begin{pmatrix}
            \bigO(\zeta^{-1}) & \bigO(\zeta^{-2})\\
            \bigO(\zeta^{-1}) & \bigO(\zeta^{-1})
        \end{pmatrix},\qquad\mbox{as $\zeta\to\infty$.}
    \end{equation}
    Since $\det \Phi\equiv 1$ we obtain from (\ref{definition: Q}) and (\ref{definition: Qhat}) that
    $\det \widehat \Psi\equiv 1$, as well. Then, it is easy to
    verify that
\[
    \widehat \Psi\sigma_3\widehat \Psi^{-1}=\begin{pmatrix}
        1+2\widehat \Psi_{12}\widehat \Psi_{21} & -2 \widehat \Psi_{11}\widehat \Psi_{12} \\
        2\widehat \Psi_{21}\widehat \Psi_{22} & -1-2\widehat \Psi_{12}\widehat \Psi_{21}
    \end{pmatrix}
    \equiv\begin{pmatrix}
        Q_{11} & -iQ_{12} \\
        -iQ_{21} & -Q_{11}
    \end{pmatrix},
\]
and hence, by (\ref{definition: N theta}),
\begin{equation}\label{NhatPsisigma3}
    N\widehat
        \Psi\sigma_3\widehat \Psi^{-1}N^{-1}=
    \begin{pmatrix}
        \frac{1}{2}(Q_{21}-Q_{12}) & -\frac{1}{2}(Q_{21}+Q_{12})-Q_{11} \\[1ex]
        \frac{1}{2}(Q_{21}+Q_{12})-Q_{11} & \frac{1}{2}(Q_{12}-Q_{21})
    \end{pmatrix}.
\end{equation}
The asymptotics of the functions $Q_{11},Q_{12}$ and $Q_{21}$ at
infinity follow from the asymptotic behavior (\ref{asymptotics
Qhat}) of $\widehat\Psi$.  We find, as $\zeta\to\infty$,
\begin{align}
    \label{R11}
    Q_{11} &=
        1+\frac{1}{2}y^2\zeta^{-2}+\bigl(yw_1-\frac{1}{2}r_1^2\bigr)\zeta^{-3}+\bigO(\zeta^{-4}), \\[1ex]
    \nonumber
    Q_{12} &=
        y\zeta^{-1}+(r_1-yh)\zeta^{-3/2}+\bigl(\frac{1}{2}yh^2-hr_1+w_1\bigr)\zeta^{-2}\\
        &\hspace{6.5cm}+\frac{1}{8}t\zeta^{-5/2}
        +u\zeta^{-3}+v\zeta^{-7/2}+\bigO(\zeta^{-4}),
    \\[1ex]
    \nonumber
    Q_{21}
    &=y\zeta^{-1}-(r_1-yh)\zeta^{-3/2}+\bigl(\frac{1}{2}yh^2-hr_1+w_1\bigr)\zeta^{-2}
    \\
    \label{R21}
        &\hspace{6.5cm}-\frac{1}{8}t\zeta^{-5/2}+u\zeta^{-3}-v\zeta^{-7/2}+\bigO(\zeta^{-4}),
\end{align}
where $t,u$ and $v$ are some functions of $x$ and $T$. Inserting
(\ref{NhatPsisigma3})--(\ref{R21}) into (\ref{proof: main theorem:
U eq1}) and using the fact that,
\[
    \frac{\partial\theta}{\partial\zeta}=\frac{1}{30}\zeta^{5/2}-\frac{1}{2}T\zeta^{1/2}+\frac{1}{2}x\zeta^{-1/2},
\]
it is straightforward to check that,
\begin{align*}
    U &= \frac{1}{240}
    \begin{pmatrix}
        a\zeta+t & 8\zeta^2+8y\zeta+b+e\zeta^{-1}\\[1ex]
        8\zeta^3-8y\zeta^2+c\zeta+d & -a\zeta-t
    \end{pmatrix}+\begin{pmatrix}
        \bigO(\zeta^{-1}) & \bigO(\zeta^{-2})\\[1ex]
        \bigO(\zeta^{-1}) & \bigO(\zeta^{-1})
    \end{pmatrix},
\end{align*}
with
\begin{align}
    \label{definition: a}
    a &= 8r_1-8yh,\\[1ex]
    \label{definition: b c}
    b &= 4y^2-120T+4yh^2-8hr_1+8w_1 ,&& c = 4y^2-120T-4yh^2+8hr_1-8w_1,\\[1ex]
    d &= 8yw_1-4r_1^2+120x+120yT-8u, && e =8yw_1-4r_1^2+120x-120yT+8u.
\end{align}
Since $U$ is entire, it contains no negative powers of $\zeta$. In
particular $e=0$, so that
\begin{equation}\label{definition: d}
    d=d+e=16yw_1-8r_1^2+240x.
\end{equation}
We now have shown that,
\begin{equation}\label{proof: main theorem: U final eq}
    U=\frac{1}{240}
    \begin{pmatrix}
        a\zeta+t & 8\zeta^2+8y\zeta+b\\[1ex]
        8\zeta^3-8y\zeta^2+c\zeta+d & -a\zeta-t
    \end{pmatrix},
\end{equation}
where $a,b$ and $c$ are given by (\ref{definition: a}) and
(\ref{definition: b c}), and where $d$ is given by
(\ref{definition: d}).

Next, we consider $W$. Observe that by (\ref{definition: Qhat}),
\begin{equation}\label{proof: main theorem: W eq1}
    W=\zeta^{-\frac{\sigma_3}{4}}N\frac{\partial\widehat\Psi}{\partial
    x}\, \widehat\Psi^{-1}N^{-1}\zeta^{\frac{\sigma_3}{4}}-\frac{\partial\theta}{\partial
    x}\zeta^{-\frac{\sigma_3}{4}}\left(N\widehat
        \Psi\sigma_3\widehat
        \Psi^{-1}N^{-1}\right)\zeta^{\frac{\sigma_3}{4}}.
\end{equation}
From (\ref{asymptotics Qhat}) we obtain
\begin{align}\label{proof: main theorem: W eq2}
    \nonumber
    \zeta^{-\frac{\sigma_3}{4}}N\frac{\partial\widehat\Psi}{\partial x}\,
    \widehat\Psi^{-1}N^{-1}\zeta^{\frac{\sigma_3}{4}}&=
    \zeta^{-\frac{\sigma_3}{4}}N\left(-h_x\sigma_3\zeta^{-1/2}+\bigO(\zeta^{-1})\right)N^{-1}\zeta^{\frac{\sigma_3}{4}}
    \\[1ex]&=\begin{pmatrix}
        0 & 0 \\
        h_x & 0
    \end{pmatrix}+\bigO(\zeta^{-1/2}),
\end{align}
where $h_x$ denotes the derivative of $h$ with respect to $x$.
Further, using (\ref{NhatPsisigma3})--(\ref{R21}) together with
the fact that $\frac{\partial\theta}{\partial x}=\zeta^{1/2}$, we
have
\begin{equation}\label{proof: main theorem: W eq3}
    -\frac{\partial\theta}{\partial
    x}\zeta^{-\frac{\sigma_3}{4}}\left(N\widehat
        \Psi\sigma_3\widehat
        \Psi^{-1}N^{-1}\right)\zeta^{\frac{\sigma_3}{4}}=
        \begin{pmatrix}
            0 & 1 \\
            \zeta-y & 0
        \end{pmatrix}+\bigO(\zeta^{-1}).
\end{equation}
Inserting (\ref{proof: main theorem: W eq2}) and (\ref{proof: main
theorem: W eq3}) into (\ref{proof: main theorem: W eq1}), and
using the fact that $W$ is entire (so that $W$ contains no
negative powers of $\zeta$) we arrive at,
\begin{equation}\label{proof: main theorem: W eq4}
    W=\begin{pmatrix}
        0 & 1 \\
        \zeta+(h_x-y) & 0
    \end{pmatrix}.
\end{equation}

We will now complete the proof by determining the functions
$a,b,c,d,t$ and $h_x$ exclusively in terms of $y$, $y_x$,
$y_{xx}$, and $y_{xxx}$, using the compatibility condition
\[
    \frac{\partial^2\Psi}{\partial\zeta \partial x}=\frac{\partial^2\Psi}{\partial x
    \partial\zeta}.
\]
This condition is equivalent to ${\displaystyle \frac{\partial
U}{\partial x} -\frac{\partial W}{\partial \zeta}+UW-WU=0}$ and
leads, after a straightforward calculation, to
\[
    C_0 \zeta^2+C_1 \zeta+C_2=0,
\]
where
\begin{align}
    C_0&=\begin{pmatrix}
        8(h_x+y) & 0 \\[1ex]
        - 8y_x-2a & 8(h_x+y)
    \end{pmatrix},\\[2ex]
    C_1&=\begin{pmatrix}
        a_x+8y(h_x-y)+b-c & 8y_x+2a \\[1ex]
        c_x-2a(h_x-y)-2t & -a_x-8y(h_x-y)-b+c
    \end{pmatrix},\\[2ex]
    C_2&=\begin{pmatrix}
        t_x+b(h_x-y)-d & b_x+2t \\[1ex]
        d_x-2t(h_x-y)-240 & -t_x-b(h_x-y)+d
    \end{pmatrix}.
\end{align}
Since $C_0=0$ we deduce that $h_x=-y$, and hence by (\ref{proof:
main theorem: W eq4}) $W$ is of the form (\ref{introduction: W}),
and that $a=-4y_x$. By (\ref{definition: a}) we then have,
\begin{equation}\label{definition: r1}
    r_1=-\frac{1}{2}y_x+yh.
\end{equation}
Further, since $C_{1,11}=0$ we then obtain from (\ref{definition:
b c}) that
\begin{equation}\label{definition: w1}
    w_1=\frac{1}{4}y_{xx}+y^2+\frac{1}{2}yh^2-\frac{1}{2}y_x h.
\end{equation}
Inserting the expressions (\ref{definition: r1}) and
(\ref{definition: w1}) for $r_1$ and $w_1$ into the expressions
(\ref{definition: a}), (\ref{definition: b c}) and
(\ref{definition: d}) for $a,b,c$ and $d$, and using the fact that
$t=-\frac{1}{2}b_x$ (since $C_{2,12}=0$) we arrive at
\begin{align}
    a &= -4y_x, && b=12y^2+2y_{xx}-120T,\\[1ex]
    c &= -4y^2-2y_{xx}-120T,&& d = 16y^3-2y_x^2+4yy_{xx}+240x,\\[1ex]
    t &= -12yy_x-y_{xxx}.
\end{align}
Inserting the latter equations into (\ref{proof: main theorem: U
final eq}) we have that $U$ is of the form (\ref{introduction: U}).
Note that the fact that $y$ satisfies the $P_I^2$ equation now
follows from $C_{2,11}=0$. This proves the first part of the
theorem.
\end{varproof}
\begin{remark}\label{remark: derivative x}
Note that, by Lemma \ref{lemma: solvability Phi} (iii), we can
safely differentiate $y$ and $h$ with respect to $x$, as we did in
the above proof.
\end{remark}

\section{Asymptotic behavior of $y(x,T)$ as $x\to\pm\infty$}
    \label{section: asymptotics  y}

In this section we will determine for fixed $T\in\mathbb R$ the
asymptotics (as $x\to\pm\infty$) of the particular solution
$y(x,T)$ of the $P_I^2$ equation with no poles on the real line as
constructed in the previous section and given by,
cf.~(\ref{definition: y}),
\begin{equation}
    y=2A_{1,11}-A_{1,12}^2.
\end{equation}
Here, $A_1$ is the matrix valued function appearing in the
asymptotic expansion (\ref{asymptotic expansion: Phi}) for $\Phi$.
So, it suffices to determine the asymptotics (as $x\to\pm\infty$)
of the first row of $A_1$ which we will do by applying the
Deift/Zhou steepest-descent method \cite{Deift,DKMVZ2,DKMVZ1,DZ1,
DZ2} to the RH problem for $\Phi$.

\subsection{Rescaling of the RH problem and deformation of the jump contour}

\begin{figure}[t]
\begin{center}
    \setlength{\unitlength}{1truemm}
    \begin{picture}(100,70)(-5,2)
        \put(50,40){\thicklines\circle*{.8}}
        \put(43,42){$z_0$}
        \put(73,40){\thicklines\circle*{.8}}
        \put(73,42){\small{0}}
        \qbezier(50,40)(47,46)(41,49.2)
        \qbezier(50,40)(47,34)(41,30.8)
        \put(41,49.2){\line(-2,1){28}}
        \put(41,30.8){\line(-2,-1){28}}
        \multiput(73,40)(-2,-1){30}{\circle*{0.1}}
        \multiput(73,40)(-2,1){30}{\circle*{0.1}}

        \put(0,40){\line(1,0){90}}
        \put(20,40){\thicklines\vector(1,0){.0001}}
        \put(83,40){\thicklines\vector(1,0){.0001}}
        \put(25,57.2){\thicklines\vector(2,-1){.0001}}
        \put(25,22.8){\thicklines\vector(2,1){.0001}}

        \put(86,43){$\hat\Gamma_1$}
        \put(7,62){$\hat\Gamma_2$}
        \put(7,43){$\hat\Gamma_3$}
        \put(7,16){$\hat\Gamma_4$}

        \put(70,55){I}
        \put(53,43){V}
        \put(53,34.5){VI}
        \put(70,25){IV}
        \put(25,48){II}
        \put(25,30){III}
    \end{picture}
    \caption{The contour $\hat\Gamma=\bigcup_{j=1}^4\hat\Gamma_j$. Note that the dotted lines are not part of the contour.}
    \label{figure: deformed contour}
\end{center}
\end{figure}
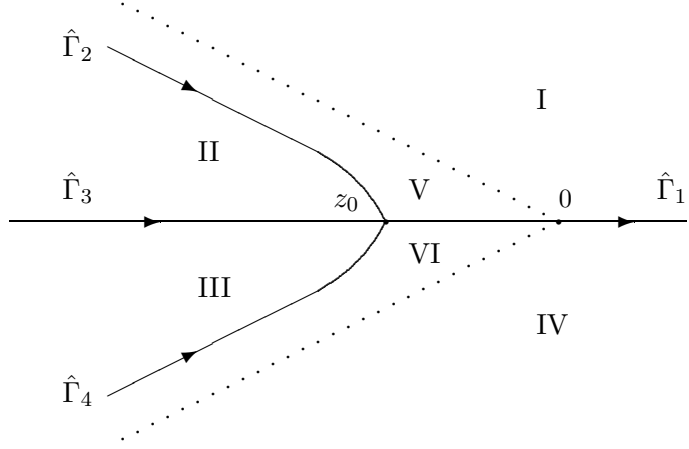

Let $z_0=z_0(x,T)\in\mathbb{R}$ (to be determined in Section
\ref{subsection: normalization}) and let
$\hat\Gamma=\bigcup_{j=1}^4\hat\Gamma_j$ be the oriented contour
through $z_0$ as shown in Figure \ref{figure: deformed contour}.
Here, the dotted lines are in fact $\Gamma_2$ and $\Gamma_4$, see
Figure \ref{figure: RHP Psi}, and are not part of the contour. The
precise form of the contour $\hat\Gamma$ (in particular of
$\hat\Gamma_2$ and $\hat\Gamma_4$) will be determined below. Now,
introduce the $2\times 2$ matrix valued function
$Y(\zeta;x,T)=Y(\zeta)$ as follows,
\begin{equation}\label{definition: T}
    Y(\zeta)\equiv\begin{cases}
        \Phi(|x|^{1/3}\zeta),&\mbox{for
        $\zeta\in\rm{I}\cup\rm{II}\cup\rm{III}\cup\rm{IV}$,}\\[2ex]
        \Phi(|x|^{1/3}\zeta)\begin{pmatrix}
            1 & 0\\
            1 & 1
        \end{pmatrix},&\mbox{for $\zeta\in\rm{V}$,}\\[3ex]
        \Phi(|x|^{1/3}\zeta)\begin{pmatrix}
            1 & 0\\
            -1 & 1
        \end{pmatrix},&\mbox{for $\zeta\in\rm{VI}$,}
    \end{cases}
\end{equation}
where $\Phi$ is the solution of the RH problem for $\Phi$, see
Section \ref{subsection: RH problem for Phi}, and where the sets
I,II,$\ldots$,VI are defined by Figure \ref{figure: deformed
contour}. Then, it is straightforward to check, using (\ref{RHP Phi:
b1})--(\ref{RHP Phi: b3}), (\ref{asymptotic expansion: Phi}) and
(\ref{definition: N theta}), that $Y$ satisfies the following
conditions.

\subsubsection*{RH problem for $Y$:}

\begin{itemize}
    \item[(a)] $Y$ is analytic in $\mathbb{C}\setminus\hat\Gamma$.
    \item[(b)] $Y$ satisfies the same jump relations on $\hat\Gamma$ as $\Phi$ does on
    $\Gamma$. Namely,
    \begin{align}
        \label{RHP T: b1}
        Y_+(\zeta)&=Y_-(\zeta)\begin{pmatrix}
            1 & 1 \\
            0 & 1
        \end{pmatrix},&\mbox{for $\zeta\in\hat\Gamma_1$,}\\[1ex]
        Y_+(\zeta)&=Y_-(\zeta)\begin{pmatrix}
            1 & 0 \\
            1 & 1
        \end{pmatrix},&\mbox{for
        $\zeta\in\hat\Gamma_2\cup\hat\Gamma_4$,}\\[1ex]
        \label{RHP T: b3}
        Y_+(\zeta)&=Y_-(\zeta)\begin{pmatrix}
            0 & 1 \\
            -1 & 0
        \end{pmatrix},&\mbox{for $\zeta\in\hat\Gamma_3$.}
    \end{align}
    \item[(c)] $Y$ has the following behavior as $\zeta\to\infty$,
    \begin{equation}\label{RHP T: c}
        Y(\zeta)\sim\left(I+\sum_{k=1}^\infty A_k
        |x|^{-k/3}\zeta^{-k}\right)\zeta^{-\frac{\sigma_3}{4}}|x|^{-\frac{\sigma_3}{12}}
        Ne^{-|x|^{7/6}\hat\theta(\zeta;x,T)\sigma_3},
    \end{equation}
    where
    \begin{equation}\label{definition: thetahat}
        \hat\theta(\zeta;x,T)=\frac{1}{105}\zeta^{7/2}-\frac{1}{3}|x|^{-2/3}T\zeta^{3/2}+\sgn(x)\zeta^{1/2}.
    \end{equation}
\end{itemize}

\subsection{Normalization of the RH problem for $Y$}
\label{subsection: normalization}

In order to normalize the RH problem for $Y$ at infinity we proceed
as Kapaev in \cite{Kapaev2}. Introduce a function
$g(\zeta;x,T)=g(\zeta)$ of the following form,
\begin{equation}\label{definition: g}
    g(\zeta)=
        c_1(\zeta-z_0)^{7/2}+c_2(\zeta-z_0)^{5/2}+c_3(\zeta-z_0)^{3/2}.
\end{equation}
where $z_0$ and the coefficients $c_1$, $c_2$, and $c_3$ are to be
chosen independent of $\zeta$ (but possibly depending on $x$ and
$T$) in such a way that
\begin{equation}
    g(\zeta)=\hat\theta(\zeta)+\bigO(\zeta^{-1/2}),
    \qquad \mbox{as $\zeta\to\infty$.}
\end{equation}

If we let $z_0=z_0(x,T)$ be the real solution of the following
third order equation (which has one real and two complex conjugate
solutions),
    \begin{equation}\label{definition: z0}
        z_0^3=-\sgn(x)48+24z_0|x|^{-2/3}T, \qquad \mbox{for $x\neq 0$,}
    \end{equation}
and if we set
\begin{equation}\label{definition: c1 c2 c3}
    c_1=\frac{1}{105}, \qquad
    c_2=\frac{1}{30}z_0, \qquad
    c_3=\frac{1}{36}z_0^2-\sgn(x)\frac{2}{3z_0},
\end{equation}
then it is straightforward to verify, using (\ref{definition:
thetahat}) and (\ref{definition: g}), that for $\zeta$ sufficiently
large,
\begin{equation}\label{asymptotics g}
    g(\zeta)=\hat\theta(\zeta)+\sum_{k=0}^\infty b_k\zeta^{-k-\frac{1}{2}},
\end{equation}
for some unimportant $b_k$'s which depend only on $x$ and $T$ and
which can be calculated explicitly. The latter equation yields
that for $\zeta$ large enough,
\begin{equation}\label{asymptotics exp theta g}
        e^{|x|^{7/6}(g(\zeta)-\hat\theta(\zeta))\sigma_3}=
        I+\sum_{k=1}^\infty d_k\sigma_3^k\zeta^{-k/2},
\end{equation}
where the coefficients $d_k$ can also be calculated explicitly.
Further, observe that by (\ref{asymptotics exp theta g}) we have
$\det(I+\sum_{k=1}^\infty d_k\sigma_3^k\zeta^{-k/2})=1$, which
yields
\begin{equation}\label{definition: d2}
    d_2=\frac{1}{2}d_1^2.
\end{equation}

\begin{figure}[t]
\begin{center}
\includegraphics[scale=0.4, angle=270]{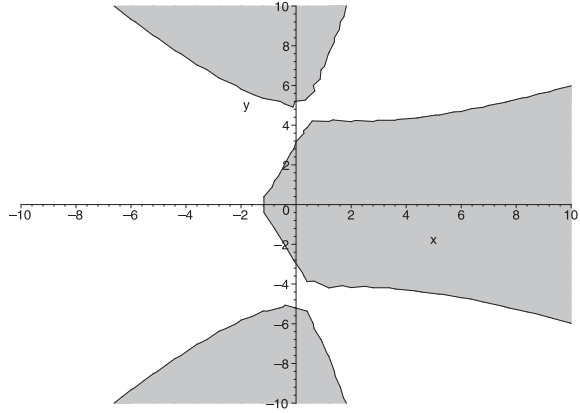}
\end{center}
\caption{Contour plot of $\Re g$ for $T=0$ and $x>0$. The shaded
areas indicate where $\Re g>0$.\label{figure: contourplot g}}
\end{figure}

Another crucial feature of the $g$-function is stated in the
following proposition, which is important for the choice of the
contour $\hat\Gamma$, and which is illustrated by Figure
\ref{figure: contourplot g}.

\begin{proposition}\label{Proposition: Re g}
    There exist constants $c>0$, $\varepsilon_0>0$ and $x_0>0$ such that for $x\geq x_0$,
    \begin{align}
    \label{Reg-1}
        &\Re g(\zeta) >c|\zeta -z_0|^{7/2}>0, &\mbox{as
        $\Arg(\zeta-z_0)=0$,}\\[1ex]
    \label{Reg-2}
        &\Re g(\zeta) <-c|\zeta -z_0|^{7/2}<0,
        &\mbox{as $\frac{6\pi}{7}-\varepsilon_0 \leq |\Arg(\zeta-z_0)|\leq \frac{6\pi}{7}+\varepsilon_0 $.}
    \end{align}
\end{proposition}

\begin{proof}
    With $\zeta=z_0+re^{i\phi}$ we have
    \begin{equation}\label{proof: proposition g: eq1}
        r^{-7/2} \Re g(\zeta) =
            c_1\cos(7\phi/2) +
            c_2 \cos(5\phi/2) r^{-1} +
            c_3\cos(3\phi/2) r^{-2},
    \end{equation}
    where by using (\ref{definition: z0}) and (\ref{definition: c1 c2 c3})
    \begin{equation}\label{proof: proposition g: eq2}
        c_1=\frac{1}{105},\qquad c_2=-\frac{1}{15}\sgn(x) 6^{1/3}+\bigO(x^{-2/3}),
        \qquad c_3=6^{-1/3}+\bigO(x^{-2/3}),
    \end{equation}
    as $|x|\to\infty$. Observe that the right hand side of (\ref{proof: proposition g: eq1})
    is a second order equation in $r^{-1}$, so that it is straightforward to check that,
    \begin{equation}
        \label{proof: proposition g: eq3}
        \min (r^{-7/6}\Re
        g(\zeta)) =c_1-\frac{c_2^2}{4c_3}=\frac{1}{350}+\bigO(x^{-2/3}),\qquad\mbox{as
        $\phi=0$,}
    \end{equation}
    which yields already (\ref{Reg-1}), and that
    \begin{equation}
        \label{proof: proposition g: eq4}
        \max (r^{-7/6}\Re
        g(\zeta))=c_1\cos(7\phi/2)-\frac{c_2^2}{4c_3}\frac{\cos^2(5\phi/2)}{\cos(3\phi/2)},
        \qquad\mbox{as $\pi/3<|\phi|<\pi$.}
    \end{equation}
    Further, since
    \[\cos(7\phi/2)=-1,\qquad  -\frac{\cos^2(5\phi/2)}{\cos(3\phi/2)}<
        1.31,\qquad\mbox{as $\phi=\frac{6\pi}{7}$,}
    \]
    there exists, by continuity in $\phi$, a constant $\varepsilon_0>0$
    sufficiently small such that the following estimates hold,
    \[
        \cos(7\phi/2)<-0.99,\qquad
        -\frac{\cos^2(5\phi/2)}{\cos(3\phi/2)}<
        1.31,\qquad\mbox{as $\frac{6\pi}{7}-
        \varepsilon_0\leq|\phi|\leq\frac{6\pi}{7}+\varepsilon_0$.}
    \]
    This implies by (\ref{proof: proposition g: eq4}) and
    (\ref{proof: proposition g: eq2}) that
    \begin{multline}
        \label{proof: proposition g: eq5}
        \max(r^{-7/6}\Re
        g(\zeta))< -0.99
        c_1+1.31\frac{c_2^2}{4c_3}<-0.00069+\bigO(x^{-2/3}),
        \\[1ex] \mbox{as $\frac{6\pi}{7}-\varepsilon_0
        \leq|\phi|\leq\frac{6\pi}{7}+\varepsilon_0$,}
    \end{multline}
    which proves (\ref{Reg-2}).
\end{proof}

\begin{remark}
    Recall that the contour $\hat\Gamma$ (in particular $\hat\Gamma_2$ and
    $\hat\Gamma_4$) is not yet explicitly defined. For now, we
    choose $\hat\Gamma_2$ and $\hat\Gamma_4$ to lie in the sectors
    where (\ref{Reg-2}) holds.
\end{remark}

\medskip

We are now ready to normalize the RH problem for $Y$ at infinity.
Let $S(\zeta;x,T)=S(\zeta)$ be the following $2\times 2$ matrix
valued function,
\begin{equation}
    S(\zeta)=
    \begin{pmatrix}
        1 & 0\\
        d_1 |x|^{1/6} & 1
    \end{pmatrix}
    Y(\zeta)e^{|x|^{7/6}g(\zeta)\sigma_3},\qquad\mbox{for $\zeta\in\mathbb{C}\setminus\hat\Gamma$,}
\end{equation}
where $Y$, $g$ and $d_1$ are given by (\ref{definition: T}),
(\ref{definition: g}) and (\ref{asymptotics exp theta g}),
respectively. It is then straightforward to check, using (\ref{RHP
T: b1})--(\ref{RHP T: b3}), using the fact that
$g_+(\zeta)+g_-(\zeta)=0$ for $\zeta\in(-\infty,z_0)$, and using
(\ref{RHP T: c}), (\ref{asymptotics exp theta g}), (\ref{definition:
N theta}) and (\ref{definition: d2}), that $S$ satisfies the
following conditions.

\subsubsection*{RH problem for $S$:}
\begin{itemize}
    \item[(a)] $S$ is analytic in $\mathbb{C}\setminus\hat\Gamma$.
    \item[(b)] $S_+(\zeta)=S_-(\zeta)v_S(\zeta)$ for $\zeta\in\hat\Gamma$, where $v_S$ is given
    by,
    \begin{equation}\label{RHP S: b}
        v_S(\zeta)=
        \begin{cases}
        \begin{pmatrix}
            1 & e^{-2|x|^{7/6}g(\zeta)}\\
            0 & 1
        \end{pmatrix},&\mbox{for $\zeta\in\hat\Gamma_1$,}\\[3ex]
        \begin{pmatrix}
            1 & 0\\
            e^{2|x|^{7/6}g(\zeta)} & 1
        \end{pmatrix},& \mbox{for
        $\zeta\in\hat\Gamma_2\cup\hat\Gamma_4$,}\\[3ex]
        \begin{pmatrix}
            0 & 1\\
            -1 & 0
        \end{pmatrix},&\mbox{for $\zeta\in\hat\Gamma_3$.}
        \end{cases}
    \end{equation}
    \item[(c)] $S$ has the following behavior as $\zeta\to\infty$,
    \begin{multline}\label{RHP S: c}
        S(\zeta)=
        \left[I+
            \begin{pmatrix}
                1 & 0 \\
                d_1 |x|^{1/6} & 1
            \end{pmatrix} A_1
            \begin{pmatrix}
                1 & 0 \\
                -d_1 |x|^{1/6} & 1
            \end{pmatrix}|x|^{-1/3}\zeta^{-1}
            \right.\\[2ex]
            \left.+\,
            \begin{pmatrix}
                \frac{1}{2}d_1^2 & -d_1 |x|^{-1/6} \\
                * & *
            \end{pmatrix}\zeta^{-1}+\bigO(\zeta^{-2})\right]
            \zeta^{-\frac{\sigma_3}{4}}|x|^{-\frac{\sigma_3}{12}} N,
    \end{multline}
    where the *'s denote unimportant functions depending only on
    $x$ and $T$.
\end{itemize}

\begin{remark}\label{remark: exponential decay}
    Note that by Proposition \ref{Proposition: Re g} the jump matrix $v_S$ on
    $\hat\Gamma_1,\hat\Gamma_2$ and $\hat\Gamma_4$
    converges exponentially fast (as $x\to \pm\infty$) to the identity matrix.
\end{remark}

\subsection{Parametrix for the outside region}

From Remark \ref{remark: exponential decay} we expect that the
leading order asymptotics of $\Phi$ will be determined by a matrix
valued function $P^{(\infty)}$ (which will be referred to as the
parametrix for the outside region) with jumps only on
$(-\infty,z_0)$ satisfying there the same jump relation as $S$ does.
Let
\begin{equation}\label{definition: Pinfty}
    P^{(\infty)}(\zeta)=
    |x|^{-\frac{\sigma_3}{12}}(\zeta-z_0)^{-\frac{\sigma_3}{4}}N,\qquad\mbox{for
    $\zeta\in\mathbb{C}\setminus(-\infty,z_0]$.}
\end{equation}
Then, using (\ref{definition: N theta}) and the fact that
$(\zeta-z_0)_-^{\frac{\sigma_3}{4}}(\zeta-z_0)_+^{-\frac{\sigma_3}{4}}=e^{-\frac{\pi
i}{2}\sigma_3}$ for $\zeta\in(-\infty,z_0)$, we obtain that
\begin{align}
    \nonumber
    P^{(\infty)}_+(\zeta)&=P^{(\infty)}_-(\zeta)N^{-1}(\zeta-z_0)_-^{\frac{\sigma_3}{4}}
    (\zeta-z_0)_+^{-\frac{\sigma_3}{4}}N\\[1ex]
    &=P^{(\infty)}_-(z)\begin{pmatrix}
        0 & 1 \\
        -1 & 0
    \end{pmatrix},\qquad \mbox{for $\zeta\in(-\infty,z_0)$.}
\end{align}
Before we can do the final transformation $S\mapsto R$ we need to do
a local analysis near $z_0$ since the jump matrices for $S$ and
$P^{(\infty)}$ are not uniformly close to each other in the
neighborhood of $z_0$.

\subsection{Parametrix near $z_0$}

In this subsection, we construct the parametrix near $z_0$. We
surround the fixed point $\hat z_0$, see (\ref{remark: asymptotics
z0}), by a disk $U_\delta=\{z\in\mathbb{C} : |z-\hat z_0|<\delta\}$
with radius $\delta>0$ (sufficiently small and which will be
determined in Proposition \ref{proposition: conformal map} below as
part of the problem) and we seek a $2\times 2$ matrix valued
function $P(\zeta;x,T)=P(\zeta)$ satisfying the following
conditions.

\subsubsection*{RH problem for $P$:}
\begin{itemize}
    \item[(a)] $P$ is analytic in $U_\delta\setminus\hat\Gamma$.
    \item[(b)] $P_+(\zeta)=P_-(\zeta)v_S(\zeta)$ for $\zeta\in \hat\Gamma\cap U_\delta$, where
    $v_S$ is the jump matrix for $S$ given by (\ref{RHP S: b}).
    \item[(c)] $P(\zeta)P^{(\infty)}(\zeta)^{-1}=I+\bigO(x^{-1})$, \qquad as $x\to \pm\infty$,
    uniformly for $\zeta\in\partial U_\delta$.
\end{itemize}

We start with constructing a matrix valued function satisfying
conditions (a) and (b) of the RH problem. This is based upon the
auxiliary RH problem for $M$ with jumps on the contour
$\Gamma^\sigma$, see Section \ref{subsection: solvability RH problem
for Phi}. The idea is that, by (\ref{RHP M: b1})--(\ref{RHP M: b3}),
the matrix valued function $M(|x|^{7/9}f(z))$ will satisfy
conditions (a) and (b) of the RH problem for $P$ if we have
appropriate biholomorphic maps $f$ on $U_\delta$ which satisfy the
following proposition.

\begin{proposition}\label{proposition: conformal map}
    There exists $x_1\geq x_0>0$ and $\delta>0$ such that for all $|x|\geq
    x_1$ there are biholomorphic maps $f=f(\cdot\,;x,T)$ on
    $U_\delta$ satisfying the following conditions.
    \begin{itemize}
        \item[1.] There exists a constant $c_0$ such that for all
        $\zeta\in U_\delta$ and $|x|\geq x_1$ the derivative of $f$ can be
        estimated by: $c_0<|f'(\zeta)|<1/c_0$ and $|\arg
        f'(\zeta)|<\varepsilon_0$ with $\varepsilon_0$ defined in Proposition \rm{\ref{Proposition: Re g}}.
        \item[2.]
        $f(U_\delta\cap\mathbb{R})=f(U_\delta)\cap\mathbb{R}$ and $f(U_\delta\cap\mathbb{C}_\pm)=f(U_\delta)
        \cap\mathbb{C}_\pm$.
        \item[3.] $\frac{2}{3}f(\zeta)^{3/2}=g(\zeta)$ for
        $\zeta\in U_\delta\setminus(-\infty,z_0]$.
    \end{itemize}
\end{proposition}

\begin{proof}
    One can verify, using (\ref{proof: proposition g: eq2}), that there exists $x_1\geq x_0>0$ sufficiently large
    and $\delta>0$ sufficiently small, such that for all $|x|\geq x_1$
    the function $f(\zeta;x,T)=f(\zeta)$ defined by
    \begin{align}\label{definition: f}
        \nonumber
        f(\zeta)
        &=\left(\frac{3}{2}c_3+\frac{3}{2}c_1(\zeta-z_0)^2+\frac{3}{2}c_2(\zeta-z_0)\right)^{2/3}(\zeta-z_0)\\[1ex]
        &=\left(\frac{3}{2}\frac{g(\zeta)}{(\zeta-z_0)^{3/2}}\right)^{2/3}(\zeta-z_0),
    \end{align}
    is analytic for $\zeta\in U_\delta$, and that $f$ is uniformly
    (in $x$ and $\zeta$) bounded in $U_\delta$. By Cauchy's theorem
    for derivatives we then also have that $f''$ is uniformly (in
    $x$ and $\zeta$) bounded in $U_\delta$ for a smaller $\delta$.
    Then, there exists a constant $C>0$ such that
    \[
        |f'(\zeta)-f'(z_0)|=\left|\int_{z_0}^\zeta
        f''(s)ds\right|\leq C |\zeta-z_0|,\qquad\mbox{for all $|x|\geq x_1$ and $\zeta\in U_\delta$.}
    \]
    Since, by (\ref{proof: proposition g: eq2}), $f'(z_0)=(\frac{3}{2}c_3)^{2/3}\geq
    const>0$ for $|x|$ large enough, this yields
    that for all $|x|\geq x_1$ (for a possible larger $x_1$) the functions
    $f$ are injective and hence biholomorphic in $U_\delta$ (for a
    possible smaller $\delta$) and that they satisy part 1 of the
    proposition.

    The second part follows from the first part (for a possible
    smaller $\delta$). The last part follows from the second part
    and from (\ref{definition: f}).
\end{proof}

Now, let $|x|\geq x_1$ and $\sigma\in(\frac{\pi}{3},\pi)$ (we will
specify our choice of $\sigma$ below), and recall that the contour
$\hat\Gamma$ is not yet explicitly defined. We suppose that
$\hat\Gamma$ is defined in $U_\delta$ as the pre-image of
$\Gamma^\sigma\cap f(U_\delta)$ under the map $f$ (so $\hat\Gamma$
depends on the parameters $x$ and $\sigma$), where
$\Gamma^\sigma=\cup_{j=1}^4\Gamma_j^\sigma$ is the jump contour for
$M$, as defined by (\ref{contourgammasigma}).
 Then, we
immediately have, by (\ref{RHP M: b1})-(\ref{RHP M: b3}) and part 3
of Proposition \ref{proposition: conformal map}, that
$M(|x|^{7/9}f(\zeta))$ satisfies conditions (a) and (b) of the RH
problem for $P$. Moreover, for any invertible analytic matrix valued
function $E$ in $U_\delta$, one has that
\begin{equation}\label{definition: P}
    P(\zeta)=E(\zeta)M(|x|^{7/9}f(\zeta)),\qquad\mbox{for $\zeta\in U_\delta\setminus\hat\Gamma$,}
\end{equation}
satisfies also conditions (a) and (b) of the RH problem for $P$. We
need $E$ to be such that the matching condition (c) is satisfied as
well. Let
\begin{equation}\label{definition: E}
    E(\zeta)=
    |x|^{-\frac{\sigma_3}{12}}(\zeta-z_0)^{-\frac{\sigma_3}{4}}(|x|^{7/9}f(\zeta))^{\frac{\sigma_3}{4}},
\end{equation}
which of course is an invertible analytic matrix valued function in
$U_\delta$. Then, using (\ref{RHP M: c}), (\ref{definition: Bk}) and
(\ref{definition: Pinfty}) we have,
\begin{equation}\label{matching}
    P(\zeta)P^{(\infty)}(\zeta)^{-1}=I+\Delta_1 |x|^{-1}
    +\Delta_2 |x|^{-4/3}+\bigO\left(|x|^{-7/3}\right),
\end{equation}
as $x\to \pm\infty$ uniformly for $\zeta\in\partial U_\delta$ and
$\sigma$ in compact subsets of $(\frac{\pi}{3},\pi)$, where
$\Delta_1$ and $\Delta_2$ are given by
\begin{equation}\label{definition: Delta1}
    \Delta_1=\frac{1}{f(\zeta)}
    \left(\frac{\zeta-z_0}{f(\zeta)}\right)^{1/2}
    \begin{pmatrix}
        0 & 0 \\
        t_1 & 0
    \end{pmatrix},\qquad \Delta_2=\frac{1}{f(\zeta)^2}
    \left(\frac{\zeta-z_0}{f(\zeta)}\right)^{-1/2}
    \begin{pmatrix}
        0 & s_1 \\
        0 & 0
    \end{pmatrix},
\end{equation}
and where $t_1$ and $s_1$ are unimportant constants given by
(\ref{definition: sk tk}). We then have shown that $P$ defined by
(\ref{definition: P}) satisfies the conditions of the RH problem
for $P$. This ends the construction of the parametrix near $z_0$.

\subsection{Final transformation}
\label{subsection: final transformation}

We will now perform the final transformation. Recall that the
contour $\hat\Gamma$ is still not yet explicitly defined. We will
now define it in terms of the (sufficiently large) parameter $x$.

Consider the fixed point $\hat z_0+\delta e^{\frac{6\pi i}{7}}$
(which depends only on $\sgn(x)$) on $\partial U_\delta$. Since
$z_0\to\hat z_0$ as $x\to \pm\infty$, see Remark \ref{remark:
asymptotics z0}, there exists $x_2\geq x_1$ sufficiently large such
that for all $|x|\geq x_2$,
\[
    \frac{6\pi}{7}-\varepsilon_0<\arg(\hat z_0+\delta e^{\frac{6\pi
    i}{7}}-z_0)<\frac{6\pi}{7}+\varepsilon_0,
\]
where $\varepsilon_0$ is defined in Proposition \ref{Proposition: Re
g}. From Proposition \ref{proposition: conformal map} we then know
that for $|x|\geq x_2$ there exists $\sigma=\sigma(x)\in
(\frac{6\pi}{7}-2\varepsilon_0,\frac{6\pi}{7}+2\varepsilon_0)$ such
that $f^{-1}(\Gamma^\sigma_2)\cap\partial U_\delta=\{\hat z_0+\delta
e^{\frac{6\pi i}{7}}\}$. By the symmetry
$\overline{f(\zeta)}=f(\overline{\zeta})$ we then also have
$f^{-1}(\Gamma^\sigma_4)\cap\partial U_\delta=\{\hat z_0+\delta
e^{-\frac{6\pi i}{7}}\}$. We now define $\hat\Gamma$ in $U_\delta$
(for $|x|\geq x_2)$ as the inverse $f$-image of the contour
$\Gamma^\sigma$. Outside $U_\delta$, we take
$\hat\Gamma_1\cup\hat\Gamma_3=\mathbb R$, $\hat\Gamma_2=\{\hat
z_0+te^{6\pi i/7}:t\geq \delta\}$, and $\hat\Gamma_4=\{\hat z_0
+te^{-6\pi i/7}:t\geq \delta\}.$ Note that by Proposition
\ref{Proposition: Re g},
\begin{align}
    \label{Reg1bis}
    &\Re g(\zeta)>c|\zeta-z_0|^{7/2}&\mbox{for $\zeta\in\hat\Gamma_1\setminus U_\delta$,}\\
    \label{Reg2bis}
    &\Re g(\zeta)<-c|\zeta-z_0|^{7/2}&\mbox{for
$\zeta\in(\hat\Gamma_2\cup\hat\Gamma_4)\setminus U_\delta$.}
\end{align}

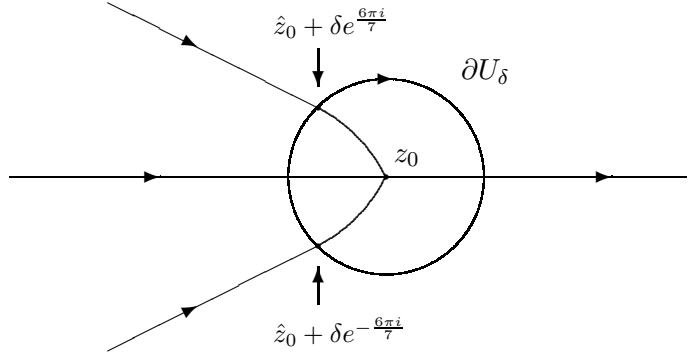
\begin{figure}[t]
\begin{center}
    \setlength{\unitlength}{1truemm}
    \begin{picture}(100,55)(-5,10)
        \cCircle(50,40){13}[f]
        \put(50,40){\thicklines\circle*{.8}}
        \put(51,42){$z_0$}
        \put(60,53){$\partial U_\delta$}
        \qbezier(50,40)(47,46)(41,49.2)
        \qbezier(50,40)(47,34)(41,30.8)
        \put(41,49.2){\line(-2,1){28}}
        \put(41,30.8){\line(-2,-1){28}}
        \put(0,40){\line(1,0){90}}
        \put(41,49.2){\thicklines\circle*{.8}}
        \put(41,57){\thicklines\vector(0,-1){5}}
        \put(35,59){\small{$\hat z_0+\delta e^{\frac{6\pi i}{7}}$}}
        \put(41,30.8){\thicklines\circle*{.8}}
        \put(41,23){\thicklines\vector(0,1){5}}
        \put(35,18){\small{$\hat z_0+\delta e^{-\frac{6\pi i}{7}}$}}
        \put(20,40){\thicklines\vector(1,0){.0001}}
        \put(80,40){\thicklines\vector(1,0){.0001}}
        \put(25,57.2){\thicklines\vector(2,-1){.0001}}
        \put(25,22.8){\thicklines\vector(2,1){.0001}}
        \put(51,53){\thicklines\vector(1,0){.0001}}
    \end{picture}
    \caption{The contour $\Gamma_R=\hat\Gamma_R\cup\partial U_\delta$.
    The part of $\Gamma_R$ inside $U_\delta$ depends
    on $x$. The rest of $\Gamma_R$ is independent of $x$.}
    \label{figure: system of contours R}
\end{center}
\end{figure}

Further define a contour $\Gamma_R$ as $\Gamma_R=\hat\Gamma\cup
\partial U_\delta$. This leads to Figure \ref{figure: system of contours R}. Note that
$\Gamma_R\cap U_\delta$ depends on $x$. However, the part of
$\Gamma_R$ outside $U_\delta$ is independent of $x$.

\medskip

Now, we are ready to do the final transformation $S\mapsto R$.
Define a $2\times 2$ matrix valued function $R(\zeta;x,T)=R(\zeta)$
for $\zeta\in\mathbb{C}\setminus\Gamma_R$ as
\begin{equation}\label{definition: R}
    R(\zeta)=
    \begin{cases}
        S(\zeta)P(\zeta)^{-1},&\mbox{for $\zeta\in U_\delta\setminus\Gamma_R$,} \\
        S(\zeta)P^{(\infty)}(\zeta)^{-1},&\mbox{for $\zeta$ elsewhere,}
    \end{cases}
\end{equation}
where $P$ is the parametrix near $z_0$ given by (\ref{definition:
P}), $P^{(\infty)}$ is the parametrix for the outside region given
by (\ref{definition: Pinfty}), and $S$ is the solution of the RH
problem for $S$.

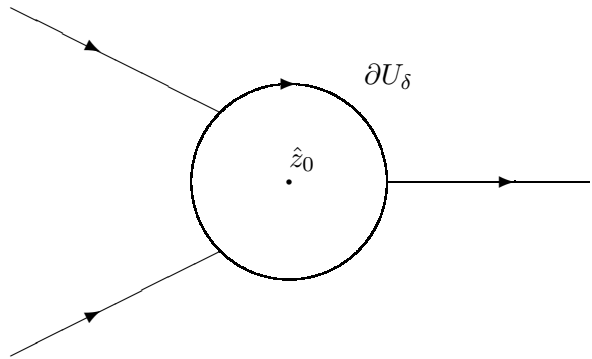
\begin{figure}[t]
\begin{center}
    \setlength{\unitlength}{1truemm}
    \begin{picture}(100,55)(-5,10)
        \cCircle(50,40){13}[f]
        \put(50,40){\thicklines\circle*{.8}}
        \put(50,42){$\hat z_0$}
        \put(60,53){$\partial U_\delta$}
        \put(41,49.2){\line(-2,1){28}}
        \put(41,30.8){\line(-2,-1){28}}
        \put(63,40){\line(1,0){27}}
        \put(80,40){\thicklines\vector(1,0){.0001}}
        \put(25,57.2){\thicklines\vector(2,-1){.0001}}
        \put(25,22.8){\thicklines\vector(2,1){.0001}}
        \put(51,53){\thicklines\vector(1,0){.0001}}
    \end{picture}
    \caption{The reduced system of contours $\hat\Gamma_R$ independent of $x$.}
    \label{figure: reduced system of contours R}
\end{center}
\end{figure}

By definition, $R$ has jumps on the contour $\Gamma_R$. However, $S$
and $P$ have the same jumps on $\Gamma_R\cap U_\delta$. Further, $S$
and $P^{(\infty)}$ satisfy the same jump relation on $(-\infty,\hat
z_0-\delta)$. This yields that $R$ has only jumps on the reduced
system of contours $\hat\Gamma_R$ (which is independent of $x$),
shown in Figure \ref{figure: reduced system of contours R}.

Using (\ref{definition: R}), (\ref{RHP S: c}) and (\ref{definition:
Pinfty}) one can now show that $R$ is a solution of the following RH
problem on the contour $\hat\Gamma_R$.

\subsubsection*{RH problem for $R$:}

\begin{itemize}
    \item[(a)] $R$ is analytic in $\mathbb{C}\setminus\hat\Gamma_R$.
    \item[(b)] $R_+(\zeta)=R_-(\zeta)v_R(\zeta)$ for
    $\zeta\in\hat\Gamma_R$, with $v_R$ given by
    \begin{align}\label{definition: vR 1}
        v_R(\zeta) &=
            P^{(\infty)}(\zeta)v_S(\zeta)P^{(\infty)}(\zeta)^{-1},
            & \mbox{for $\zeta\in\hat\Gamma_R\setminus \partial
            U_\delta$.}
        \\[1ex]
        v_R(\zeta) &=
            P(\zeta)P^{(\infty)}(\zeta)^{-1},
            & \mbox{for $\zeta\in\partial U_\delta$.}
    \end{align}
    \item[(c)] $R(\zeta)=I+\bigO(\zeta^{-1})$ as
    $\zeta\to\infty$.
\end{itemize}

\begin{remark}
    Observe that by (\ref{matching}), (\ref{Reg1bis}) and (\ref{Reg2bis}) we
    have as $x\to \pm\infty$,
    \begin{equation}\label{estimate:vR}
        v_R(\zeta)=
        \begin{cases}
            I+\Delta_1 |x|^{-1}+\Delta_2 |x|^{-4/3}+\bigO(|x|^{-7/3}), &
            \mbox{uniformly for $\zeta\in \partial U_\delta$,}
        \\
            I+\bigO(e^{-c |x|^{7/6}|\zeta -z_0|^{7/2}}) & \mbox{uniformly for $\zeta\in\hat\Gamma_R\setminus\partial U_\delta$,}
    \end{cases}
    \end{equation}
    for some constant $\gamma>0$, and where $\Delta_1$ and $\Delta_2$ are given by
    (\ref{definition: Delta1}). As in \cite{Deift, DKMVZ2, DKMVZ1},
    this yields that $R$ itself is uniformly close to the identity
    matrix,
    \[
        R(\zeta)=I+\bigO(x^{-1}),\qquad\mbox{as $x\to \pm\infty$,
        uniformly for $\zeta\in\mathbb{C}\setminus\hat\Gamma_R$.}
    \]
\end{remark}

\begin{remark}
    Since $R(\zeta)=S(\zeta)P^{(\infty)}(\zeta)^{-1}$ for
    $\zeta$ large one can use (\ref{RHP S: c}), (\ref{definition: Pinfty}), and the fact
    that
    $(\zeta-z_0)^{\frac{\sigma_3}{4}}=\zeta^{\frac{\sigma_3}{4}}
    \left[I-\frac{1}{4}z_0\sigma_3\zeta^{-1}+\bigO(\zeta^{-2})\right]$
    as $\zeta\to\infty$, to strengthen condition (c) of the RH problem for $R$ to
    \begin{equation}
    R(\zeta)=I+\frac{R_1}{\zeta}+\bigO(\zeta^{-2}),\qquad\mbox{as
        $\zeta\to\infty$,}
    \end{equation}
    where $R_1$ is a $2\times 2$ matrix valued function depending on
    $x$ and $T$ with $(1,1)$ and $(1,2)$ entries
    given by,
    \begin{align}
    \label{definition: R111}
        R_{1,11} &= -\frac{z_0}{4}+\frac{1}{2}d_1^2+|x|^{-1/3}A_{1,11}-
        d_1 |x|^{-1/6}A_{1,12}, \\[1ex]
    \label{definition: R112}
        R_{1,12} &= -d_1 |x|^{-1/6}+|x|^{-1/3}A_{1,12}.
    \end{align}
    From (\ref{estimate:vR}) it follows, as in \cite{DKMVZ2}, that
    \[R_1=-\Res(\Delta_1,z_0)|x|^{-1}-\Res(\Delta_2,z_0)|x|^{-4/3}+\bigO(|x|^{-7/3}),
    \qquad\mbox{as $x\to \pm\infty$,}\]
    so that by (\ref{definition: Delta1}),
    \begin{equation}
            \label{asymptotics: R111 R112} R_{1,11} = \bigO(|x|^{-7/3}),\qquad
            R_{1,12} =
            \bigO(|x|^{-4/3}),\qquad \mbox{as $x\to \pm\infty$.}
    \end{equation}
\end{remark}

\subsection{Proof of Theorem \ref{main theorem} (ii)}

We now have all the necessary ingredients to prove the second part
of the main theorem.

\begin{varproof}\textbf{of Theorem \ref{main theorem} (ii).}
Recall that $y=2A_{1,11}-A_{1,12}^2$. Using (\ref{definition: R111})
and (\ref{definition: R112}) one can then write $y$ in terms of the
$(1,1)$ and $(1,2)$ entries of $R_1$,
\begin{align*}
    & 2 A_{1,11}=
        \frac{1}{2}z_0 |x|^{1/3}+2|x|^{1/3}R_{1,11}-d_1^2 |x|^{1/3}+2d_1 |x|^{1/6}A_{1,12},
    \\[1ex]
    & A_{1,12}^2=
        |x|^{2/3}R_{1,12}^2-d_1^2 |x|^{1/3}+2d_1 |x|^{1/6} A_{1,12},
\end{align*}
so that
\begin{equation}
    y=\frac{1}{2}z_0|x|^{1/3}+2|x|^{1/3}R_{1,11}-|x|^{2/3}R_{1,12}^2.
\end{equation}
Inserting (\ref{asymptotics: R111 R112}) into the latter equation we
obtain precisely (\ref{main theorem: eq1}). This finishes the proof
of Theorem \ref{main theorem}.
\end{varproof}

%

\section*{Acknowledgements}

We are grateful to Arno Kuijlaars for careful reading and very
useful remarks and discussions. We also like to thank Maurice
Duits for stimulating discussions.

The authors are supported by FWO research project G.0455.04, by
K.U.Leuven research grant OT/04/24, and by INTAS Research Network
NeCCA 03-51-6637. The second author is Postdoctoral Fellow of the
Fund for Scientific Research - Flanders (Belgium).

\end{document}